\documentclass[a4paper,11pt]{article}
\pdfoutput=1 % if your are submitting a pdflatex (i.e. if you have
\usepackage{jcappub} % for details on the use of the package, please
\usepackage[T1]{fontenc} % if needed

\title{Bounce Inflation Cosmology with Standard Model Higgs Boson}

\author[a]{Youping Wan,}
\author[b,c]{Taotao Qiu,}
\author[a]{Fa Peng Huang,}
\author[d]{Yi-Fu Cai,}
\author[e]{Hong Li,}
\author[a]{and Xinmin Zhang}

\affiliation[a]{Theoretical Physics Division, Institute of High Energy Physics,
Chinese Academy of Sciences,\\P.O.Box 918-4, Beijing 100049, P.R.China}
\affiliation[b]{Institute of Astrophysics, Central China Normal University,\\Wuhan 430079, P.R.China}
\affiliation[c]{State Key Laboratory of Theoretical Physics, Institute
of Theoretical Physics, Chinese Academy of Sciences,\\Beijing 100190, P.R.China}
\affiliation[d]{CAS Key Laboratory for Researches in Galaxies and Cosmology,
Department of Astronomy, University of Science and Technology of China,
Chinese Academy of Sciences,\\Hefei, Anhui 230026, China}
\affiliation[e]{Key Laboratory of Particle Astrophysics, Institute of High Energy Physics,
Chinese Academy of Science,\\P. O. Box 918-3, Beijing 100049, P. R. China}

\emailAdd{wanyp@ihep.ac.cn}
\emailAdd{qiutt@mail.ccnu.edu.cn}
\emailAdd{huangfp@ihep.ac.cn}
\emailAdd{yifucai@ustc.edu.cn}
\emailAdd{hongli@ihep.ac.cn}
\emailAdd{xmzhang@ihep.ac.cn}

\abstract{
It is of great interest to connect cosmology in the early universe to the Standard
Model of particle physics. In this paper, we try to construct a bounce inflation
model with the standard model Higgs boson, where the one loop correction is taken
into account in the effective potential of Higgs field. In this model, a Galileon
term has been introduced to eliminate the ghost mode when bounce happens. Moreover,
due to the fact that the Fermion loop correction can make part of the Higgs potential
negative, one naturally obtains a large equation of state(EoS) parameter  in the
contracting phase, which can eliminate the anisotropy problem. After the bounce,
the model can drive the universe into the standard higgs inflation phase, which
can generate nearly scale-invariant power spectrum.
}


\arxivnumber{1509.08772}

\begin{document}
\maketitle
\flushbottom

\section{Introduction}

It has been proven that both the Hot Big Bang (HBB) theory and inflation theory
\cite{Guth:1980zm, Linde:1981mu, Albrecht:1982wi}
(see also \cite{Starobinsky:1980te,Fang:1980wi,Sato:1980yn} for early works) suffer
from a gravitational singularity problem \cite{Hawking1970,Hawking1973,Vilenkin1994a,
Vilenkin1994b}. In these theories,
the universe has to start from a spacetime singularity, where all the observables
such as temperature, density and so on, become infinite. However, there are many
possibilities to obviate these problems, one of which is to have a "bounce" before
inflation in the early stage \cite{Novello:2008ra,Cai:2014bea, Battefeld}. In such a bouncing universe
\cite{Ayon-Beato:2015eca}, the scale factor first decreases, and then after reaching
its minimal value at the bounce point, it increases, making the universe expand as
is observed. As long as the minimal value of the scale factor is larger than 0, the
singularity will be avoided.

According to the Singularity Theorems, to construct a bounce before inflation, we
always have to violate the Null Energy Condition(NEC) if we still restrict ourselves
to General Relativity in 4D space-time \cite{Novello:2008ra,Battefeld}. The NEC can be expressed
as $T_{\mu\nu}n^\mu n^\nu=\rho+p\geq 0$ where $n^{\mu}$ is any null vector and $\rho$
and $p$ are the energy density and pressure of the universe, implying that the Equation
of State (EoS) parameter $w\equiv p/\rho$ has to be larger than $-1$. However the bounce
requires $H=0$ and $\dot H>0$ at the bouncing point, making the EoS parameter below $-1$
at least at the bounce point. Moreover, after the bounce the EoS parameter has to go back
to above $-1$, in order to connect with the observed expanding universe. This requires the
EoS parameter be cross the $w=-1$ boundary in the bounce model \cite{Cai:2007qw}.

The crossing behavior of the EoS parameter can be realized in Quintom models, which have
been proposed in the early paper of \cite{Feng:2004ad}. The Quintom model has first been
considered as a model of dark energy, and there have been Quintom models of double-field
\cite{Feng:2004ad,Guo:2004fq,Zhang:2005eg}, higher-derivative single field
\cite{Li:2005fm,Zhang:2006ck,Cai:2007gs} and many other models. Later on, Quintom has been
applied to bounce models, e.g., see investigations on double-field bounce models in
\cite{Cai:2007zv,Cai:2008qb,Cai:2008qw,Qiu:2010dk}, higher-derivative single field bounce
models in \cite{Qiu:2011cy, Easson:2011zy, Cai:2012va, Qiu:2013eoa, Koehn:2013upa, Battarra, Qiu:2015nha} and nonminimal
coupling field bounce models in \cite{Qiu:2010vk}. More comprehensive studies on Quintom
models are presented in the reviews \cite{Cai:2009zp,Qiu:2010ux}.

The phenomenology of bounce inflation scenario has been studied in \cite{Piao:2003zm,
Piao:2005ag,Piao:2003hh}.
However, one of the potential problems in realistic model building is the anisotropy
problem \cite{Kunze:1999xp,Erickson:2003zm,Xue:2010ux,Xue:2011nw}. Consider even very
small anisotropy of space-time in the
initial time of the contracting phase, they will grow proportional to $a^{-6}$ and will
in general dominate the universe's total energy density, thus the cosmological principle
will be destroyed and the space-time cannot be described by the FLRW metric, thus the
universe may collapse into a totally anisotropic one instead of bounce and enter
into an expanding one. One of the most elegant ways to avoid the anisotropy problem is
to let the total energy density dominated by some background component rather than the
anisotropies, which requires the background evolving with an EoS parameter larger than
$+1$(see Ekpyrotic models \cite{Khoury:2001wf}, or other nonsingular bounce models like
\cite{Qiu:2013eoa}).

It is known that in the Standard Model of particle physics, the only basic scalar particle that has been found in laboratory is the Higgs boson \cite{Aad:2012tfa,Chatrchyan:2012xdj}, therefore it is of great interest to ask whether the Higgs field can be connected to early universe cosmology. In the literature, Higgs field has been eagerly studied in cosmology \cite{Espinosa:2007qp, Bars:2013vba}, and in particular, see \cite{CervantesCota:1995tz, Bezrukov:2007ep, DeSimone:2008ei, Burgess:2009ea, Barbon:2009ya, Barvinsky:2009fy, Lerner:2009na, Burgess:2010zq, Hertzberg:2010dc, Germani:2010gm, Bezrukov:2010jz, Atkins:2010yg, Kamada:2010qe, Horvat:2011wr, Qiu:2011tk, Bezrukov:2013fka, Kearney:2015vba, Moss:2015fma, Huang:2013oua, Hamada:2014iga, Hamada:2014wna,Salvio:2013rja,Salvio:2015kka} for the trial to build inflation models with Higgs field, which is proved to be quite economic and predictive, and see \cite{Cai:2012qi, Kunimitsu:2012xx, Choi:2012cp, DeSimone:2012qr, DeSimone:2012gq, Cai:2013caa} for using Higgs as curvatons. In this paper, we try to use Higgs field to build a bounce inflation model.

The first step to have a bounce inflation scenario is to provide a contracting phase with large EoS parameter, in order to avoid the anisotropy problem. As can be seen below, a large EoS parameter can be obtained by a scalar field with a {\it negative} potential. For the Standard Model Higgs field, if we consider loop corrections the effective potential can be made negative at the energy range of $10^{10}-10^{12}$ GeV, but by considering the uncertainty of top quark mass and the strong coupling constant, the Higgs self-coupling constant can be negative near the Planck scale \cite{Buttazzo:2013uya, Degrassi:2012ry}. Next, in the bounce region, since the standard Higgs field itself cannot violate NEC and trigger the bounce, we introduce a Galileon \cite{Nicolis:2008in, Deffayet:2009wt, Nicolis:2009qm, Deffayet:2009mn, Deffayet:2011gz} term which contains high-derivative operator, such that the bounce can not only occur, but also without any ghost modes \cite{Qiu:2011cy, Easson:2011zy, Cai:2012va, Qiu:2013eoa, Koehn:2013upa}. Finally, the Higgs field will drive a period of inflation, which can generate a scale-invariant power spectrum to meet with the data. After inflation, it will fall into the minimum of its potential and oscillate, so as to reheat our universe. Note that in the work \cite{Brandenberger:2015nua}, we have constructed a matter bounce \cite{Cai:2013kja} cosmological model by using the same Higgs effective potential we are using here, it's
pointed out in a matter bounce universe the scale-invariant power spectrum can be generated in the contracting phase \cite{Finelli:2001sr, Wands:1998yp}.

Our paper is organized as follows: in Sec. 2 we study the bounce and inflation scenario realized by our model. In Sec. 2.1 we give a overview of the Higgs trajectory; in Sec. 2.2 we show our calculation results in the contracting phase; in Sec. 2.3 we show the cosmology near the bounce point; in Sec. 2.4 we show the slow-roll inflation we have got and the values of cosmological observables in our model. In Sec. 3 we show that our model is stable. And in the last section we make our conclusion. In the whole text, we take the reduced Planck mass scale $M_p\equiv 1/\sqrt{8\pi G_{N}}=1$, in which $G_{N}$ is the Newtonian gravitational constant, and the sign difference of the metric to be Landau-Lifshitz type $\left(+,-,-,-\right)$.

\section{The Higgs Bounce Inflation cosmology}

In this paper, we will start with the following action
\begin{eqnarray}
\label{action}
&&S_J = \int dx^4 \sqrt{-g}\left\{-\frac{M_p^2}{2}A(h)R+{\cal L}_h\right\}~, \\ \nonumber
&&{\cal L}_h = c(h)\left[hX\Box{h}+\gamma X^2\right]+X-V(h)~,
\end{eqnarray}
where $h$ is the Higgs field, the functions in the Lagrangian are:
\begin{eqnarray}
\label{functions}
&&A(h) = 1+\xi \frac{h^2}{M_p^2}~, \\ \nonumber
&&V(h) = \frac{1}{4}\lambda h^4-b \ln\left(\frac{h^2}{\Lambda^2}\right)h^4+g\frac{h^6}{\tilde{M}^2}~, \\ \nonumber
&&c(h) =
\left\{
\begin{split}
& \frac{\alpha}{M_p^4}~, ~~{\rm if} ~~|h-h_B|<\delta \\
& 0~, ~~~~~~{\rm if} ~~|h-h_B|>\delta \\
\end{split}~,
\right.~~~\gamma\equiv \frac{\beta}{\alpha}~,
\end{eqnarray}
and
\begin{eqnarray}
X \equiv \frac{1}{2}g^{\mu\nu}\frac{\partial h}{\partial x^{\mu}}\frac{\partial h}{\partial x^{\nu}} ~~
\end{eqnarray}
is the normal kinetic term.

Here several comments about the functions in (\ref{functions})
will be in order. First of all, we have introduced a non-minimal coupling function $A(h)$,
since we hope our model to inherit the success of Higgs Inflation model \cite{Bezrukov:2007ep}.
In Higgs Inflation model, the nominimal coupling between the SM Higgs field and the gravitation
field is introduced to avoid the large tensor-to-scalar ratio. Next, we suppose the Higgs
Lagrangian has a G-function term and a nonstandard $X^2$ term, which will be useful to get a bounce
even without introducing ghost modes. However, if these two operators always exist during the
whole evolution, the kinetic term of the Higgs field will become more and more important,
driving the universe into a big rip singularity \cite{Qiu:2011cy,Easson:2011zy}.
In order to solve the problem, we simply put a delta function $c(h)$ in front of them.
The delta function $c(h)$ requires those two terms only dominate the Lagrangian when
the field is near the bounce point $h_B$, with $\delta$ the width. At last, for the potential,
one can find that the first term of $V(h)$ in (\ref{functions}) is nothing but the potential
of the standard model Higgs field, where since we're talking about physics beyond GeV scale,
we neglected the terms related to its value of expected vacuum (vev). However, we add two
more terms to the potential. One is the one-loop correction from top quark, we make use
of the analytical formula of Coleman-Weinberg potential \cite{ColemanWeinberg, Sher89, RHBRMP}.
The other is the dimension-six Higgs self interaction term, which may come from many
quantum gravity theories \cite{ArkaniHamed:2008ym} and other theories \cite{Datta:1999dw}.

In this model there are 8 parameters in total. The tree level self-coupling constant $\lambda$
has been measured by particle physics experiments \cite{Aad:2012tfa,Chatrchyan:2012xdj}.
The parameter $b$ can be calculated for various particles which can give loop-corrections
to Higgs potential with the cut-off scale $\Lambda$. In SM, the largest
contribution comes from top quark loop-correction, so in this paper we simply neglect couplings
to other particles, and it is just such fermion-Higgs loop-correction makes the Higgs potential
to be negative. The parameter $\xi$ in the nonminimal coupling term
describes the coupling strength between the Higgs field and the gravitational field
\cite{Bezrukov:2007ep}, while the introduction of functions $c(h)$ will give us two more
parameters, which we denote as $\alpha$ and $\beta$ later. Finally,
the parameters $g$ and $\tilde{M}$ describe the new physics when the field approaches
Planck scale.

In our numerical calculation,we will get a bounce inflation solution by
taking the following parameters
\begin{eqnarray}
\label{parameters}
&&\lambda = 0.129~, ~~~b = \frac{3g_{Y}^4}{64\pi^2}~, ~~~g_{Y} = 0.995756~, \nonumber\\
&&\xi = 3\times 10^5~,~~~\alpha=9.6\times 10^3~,~~~\beta=3.84\times 10^4~, \nonumber\\
&&\Lambda = 2\times 10^{-2}M_p~, ~~~g = 3\times 10^{-4}~, ~~~\tilde{M} = 1 M_p~.
\end{eqnarray}

\begin{figure}[tbp]
\centering % \begin{center}/\end{center} takes some additional vertical space
\includegraphics[width=.65\textwidth]{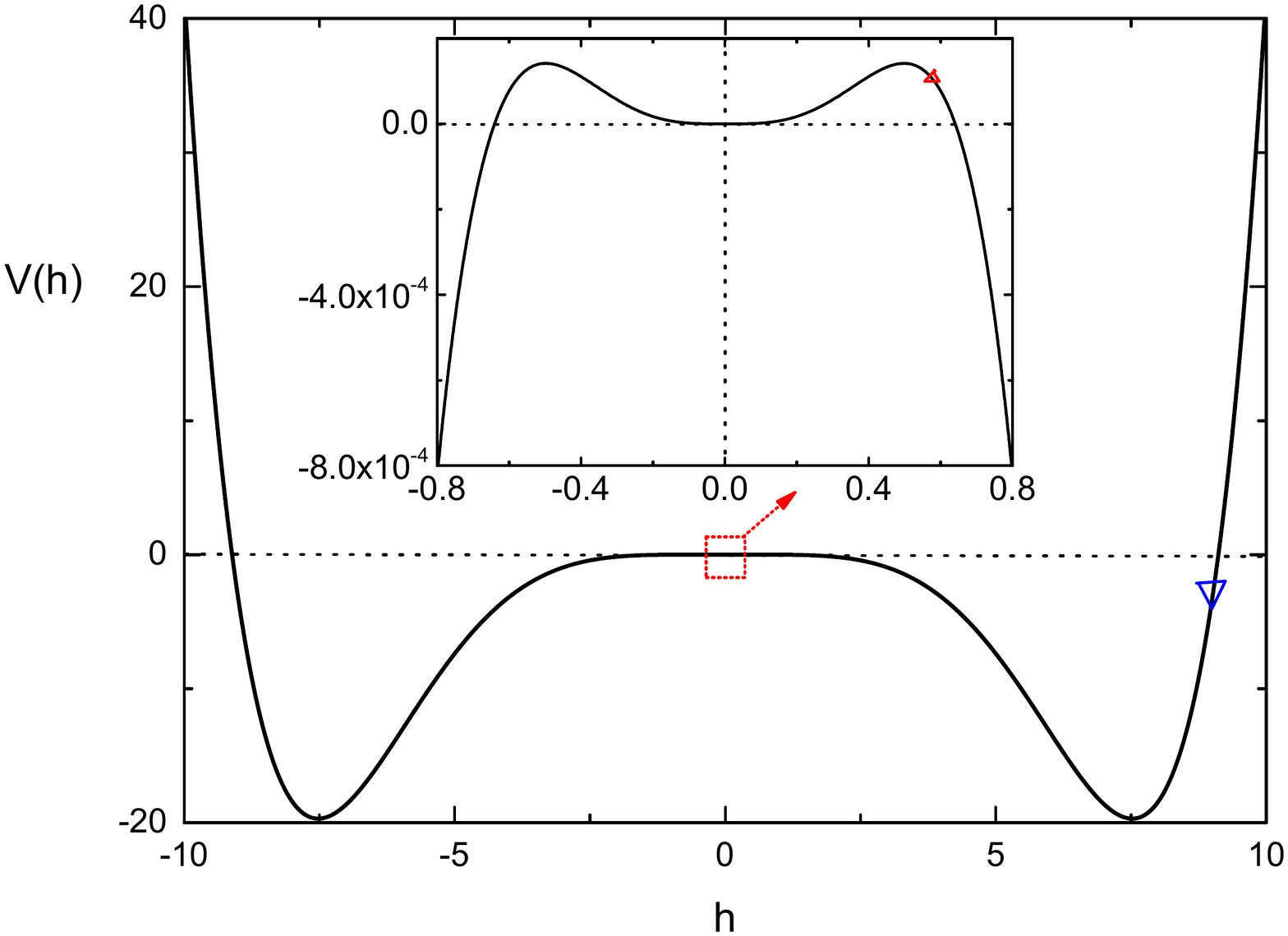}
% "\includegraphics" is very powerful; the graphicx package is already loaded
\caption{\label{fig:i} (Color plot.) The Higgs effective potential we have taken in this paper. The
blue triangle shows the initial location of $h$. There are two vacuums: one is the true vacuum
located near $h=\pm7.5 M_p$, the other is the false vacuum at $h=0$. Part of the potential is
negative due to the Fermion-loop correction. The zoomed-in plot: a closer look, the red triangle
is the bounce point. In our numerical calculation, the bounce zone where $c(h)$ becomes the
dominated part is near $0.57 M_p$, and the slow-roll inflation happen when the Higgs field
changes its value from $0.0093 M_p$ to $0.0014 M_p$.}
\label{potentialplot}
\end{figure}

In Figure \ref{potentialplot}, we illustrate the potential we have used in this paper, the blue
triangle stands for the very beginning location of $h$ and the red triangle stands for the bounce
point. We can see there are two vacuum points, one is the true vacuum near the point
$h=\pm7.5 M_p$, and the other is the false vacuum at $h=0$. Taking the $h>0$ half as an example,
there is a peak around $h=0.5 M_p$, and a zero point around $h=0.65 M_p$. As will be seen later,
the bounce point(the red triangle) is set at the middle between the peak and the zero point, i.e,
near to $h=0.57 M_p$. So after the bounce, the Higgs field needs to climb the hill to go across
the peak at $h=0.5 M_p$, and then falling down to pass through the slow-roll zone, finally
oscillates around the $h=0$.

Rather than in its original Jordan frame, it is convenient to study the model (\ref{action})
in its Einstein frame. To do this, we conduct conformal transformation to translate the
Jordan-frame action to the Einstein-frame action. The two frames are connected through
\begin{eqnarray}
\tilde{g}_{\mu\nu}=\Omega^2(h) g_{\mu\nu}~,
\end{eqnarray}
where $g_{\mu\nu}$ with and without tilde denotes metric in Einstein and Jordan frame
respectively, and here we set $\Omega^2(h)=A(h)$.

We can write down the action in Einstein
frame:
\begin{eqnarray}
\label{ActionEFrame}
S_{E}&=&\int d\tilde{x}^{4}\sqrt{-\tilde{g}}\left\{ -\frac{M_{p}^{2}}{2}\tilde{R}+\frac{6\xi^{2}}{M_{p}^{2}}h^{2}\tilde{X}A^{-2}(h)+\tilde{X}A^{-1}(h)-V(h)A^{-2}(h)\right.\nonumber\\
&&\left.+c(h)\left[\tilde{X}h\tilde{\Box}h-\frac{4\xi}{M_{p}^{2}}h^{2}\tilde{X}^{2}A^{-1}(h)+\frac{\beta}{\alpha}\tilde{X}^2\right]\right\} ~,
\end{eqnarray}
in which
\begin{eqnarray}
&&\tilde{X} = \frac{1}{2}\tilde{g}^{\mu\nu}\frac{\partial h}{\partial x^{\mu}}\frac{\partial h}{\partial x^{\nu}}~,~~\\
&&\tilde{\Box}h = \tilde{g}^{\mu\nu}\left[\frac{\partial^{2}h}{\partial x^{\mu}\partial x^{\nu}}
-\tilde{\Gamma}^{\alpha}_{\mu\nu}\frac{\partial h}{\partial x^{\alpha}}\right]~.
\end{eqnarray}

In the following parts of this paper, we'll use the definition
\begin{eqnarray}
\dot{f}\equiv\frac{d f}{d t}~,~~~f'\equiv\frac{d f}{d\tilde{t}}~
\end{eqnarray}
to denote the time derivatives. Therefore, the Friedmann equations as well as the continuity
equation of our model in Einstein frame (\ref{ActionEFrame}) can be written as:
\begin{eqnarray}\label{CosmologyEqs}
&&{\cal H}^2 = \frac{1}{3M_p^2}\rho,\\ \nonumber
&&{\cal H}' = -\frac{1}{2M_p^2}\left(\rho+p\right), \\ \nonumber
&&\rho'+3{\cal H}\left(\rho+p\right)=0,
\end{eqnarray}
in which ${\cal H}\equiv d \tilde{a}/(\tilde{a}d\tilde{t})$, and
$\tilde{a}\equiv a \Omega(h)$.

\subsection{A short overview of the Higgs trajectory}

To be more clear, we'd like to give a short overview of the Higgs trajectory.
Below we summarize according to three phases:
\begin{itemize}
\item the contracting phase. The initial conditions are taken when the universe is contracting,
the Higgs field is located at the blue triangle in Figure \ref{potentialplot},
with a leftward velocity. The Higgs field will run leftward from the blue triangle to
the red triangle.
\item the bouncing phase. Near the bounce point, the universe bounces to the expanding
phase while the Higgs field is running across the bounce point(the red triangle).
\item the expanding phase. After the universe goes into the expanding phase, the Higgs
field runs leftward further, goes across the slow-roll zone, finally oscillates around the
point of $h=0$.
\end{itemize}

\subsection{Contracting Phase}

In our model, the universe begins from a contracting phase. Due to the delta function $c(h)$, the
second line of action (\ref{ActionEFrame}) will not appear in the contracting phase, so the
Lagrangian can be written as
\begin{eqnarray}
\label{ActionEFRameConExp}
S_{E}&=&\int d\tilde{x}^{4}\sqrt{-\tilde{g}}\left[ -\frac{M_{p}^{2}}{2}\tilde{R}+\frac{6\xi^{2}}{M_{p}^{2}}h^{2}\tilde{X}A^{-2}(h)+\tilde{X}A^{-1}(h)-V(h)A^{-2}(h)\right]~,
\end{eqnarray}
and the energy and pressure of Higgs field are
\begin{eqnarray}
\label{rhocon}
\rho&=&A^{-2}(h)\left\{\left(1+\xi\frac{h^2}{M_p^2}+6\xi^2\frac{h^2}{M_p^2}\right)\frac{h'^2}{2}+V(h)\right\}~,\\
\label{prescon}
p&=&A^{-2}(h)\left\{\left(1+\xi\frac{h^2}{M_p^2}+6\xi^2\frac{h^2}{M_p^2}\right)\frac{h'^2}{2}-V(h)\right\}~,
\end{eqnarray}
while the equation of motion for the Higgs field $h$ is:
\begin{eqnarray}
\label{eomcon}
&&h''+3{\cal H}h'+\frac{1}{2}\left[-2A^{-3}(h)\left(1+\xi\frac{h^2}{M_p^2}+6\xi^2\frac{h^2}{M_p^2}\right)\frac{\partial A(h)}{\partial h}+A^{-2}(h)\left(\frac{2\xi h}{M_p^2}+\frac{12\xi^2h}{M_p^2}\right)\right]h'^2\nonumber\\
&&-2A^{-3}(h)V(h)\frac{\partial A(h)}{\partial h}+A^{-2}(h)\frac{\partial V(h)}{\partial h}=0~.
\end{eqnarray}

We set the initial conditions such that the Higgs field is located at the blue triangle in
Figure \ref{potentialplot} with a leftward velocity. Since the potential is negative
(from Figure \ref{potentialplot}), the EoS parameter turns to be larger than 1. A nonzero velocity
is required to guarantee the positivity of the total energy density, and a leftward velocity
(which is negative) can make the Higgs field climb up along the potential (whose value is
decreasing).

Since initially we are in the region where we have $\xi(h/M_p)^2\gg 1$,
here $\xi$ is set to be a large value in Eq. (\ref{parameters}), we can use the approximation:
\begin{equation}
A(h)\simeq \xi \frac{h^2}{M_p^2}~,~~~1+\xi\frac{h^2}{M_p^2}+6\xi^2\frac{h^2}{M_p^2}\simeq 6\xi^2\frac{h^2}{M_p^2}~,
\end{equation}
thus Eq.s (\ref{rhocon}), (\ref{prescon}) and (\ref{eomcon}) can be simplified as:
\begin{eqnarray}
\label{rhoprescon2}
&&\rho\simeq\left(\frac{\xi h^2}{M_p^2}\right)^{-2}\left(\frac{3\xi^2h^2}{M_p^2}h'^2+V(h)\right)~,~~~p\simeq\left(\frac{\xi h^2}{M_p^2}\right)^{-2}\left(\frac{3\xi^2h^2}{M_p^2}h'^2-V(h)\right)~,\\
\label{eomcon2}
&&h''+3{\cal H}h'+\frac{6M_p^2h'^2}{h^3}+\frac{2M_p^4b}{\xi^2h}+\frac{2gM_p^4h}{\xi^2\tilde{M}^6}=0~.
\end{eqnarray}
Since $V(h)$ is negative at the initial point, the only constraint on the initial
conditions is from the requirement of the positivity of energy density $\rho$. From Eq.
(\ref{rhoprescon2}) and the expression of $V(h)$ in (\ref{functions}), this requires
\begin{equation}
h_i'<-\frac{M_p}{\sqrt{3}\xi}\sqrt{\left[b\ln\left(\frac{h_i^2}{\Lambda^2}\right)-\frac{\lambda}{4}\right]h_i^2-g\frac{h_i^4}{\tilde{M}^2}}~.
\end{equation}

\begin{figure}[tbp]
\centering % \begin{center}/\end{center} takes some additional vertical space
\includegraphics[width=.45\textwidth]{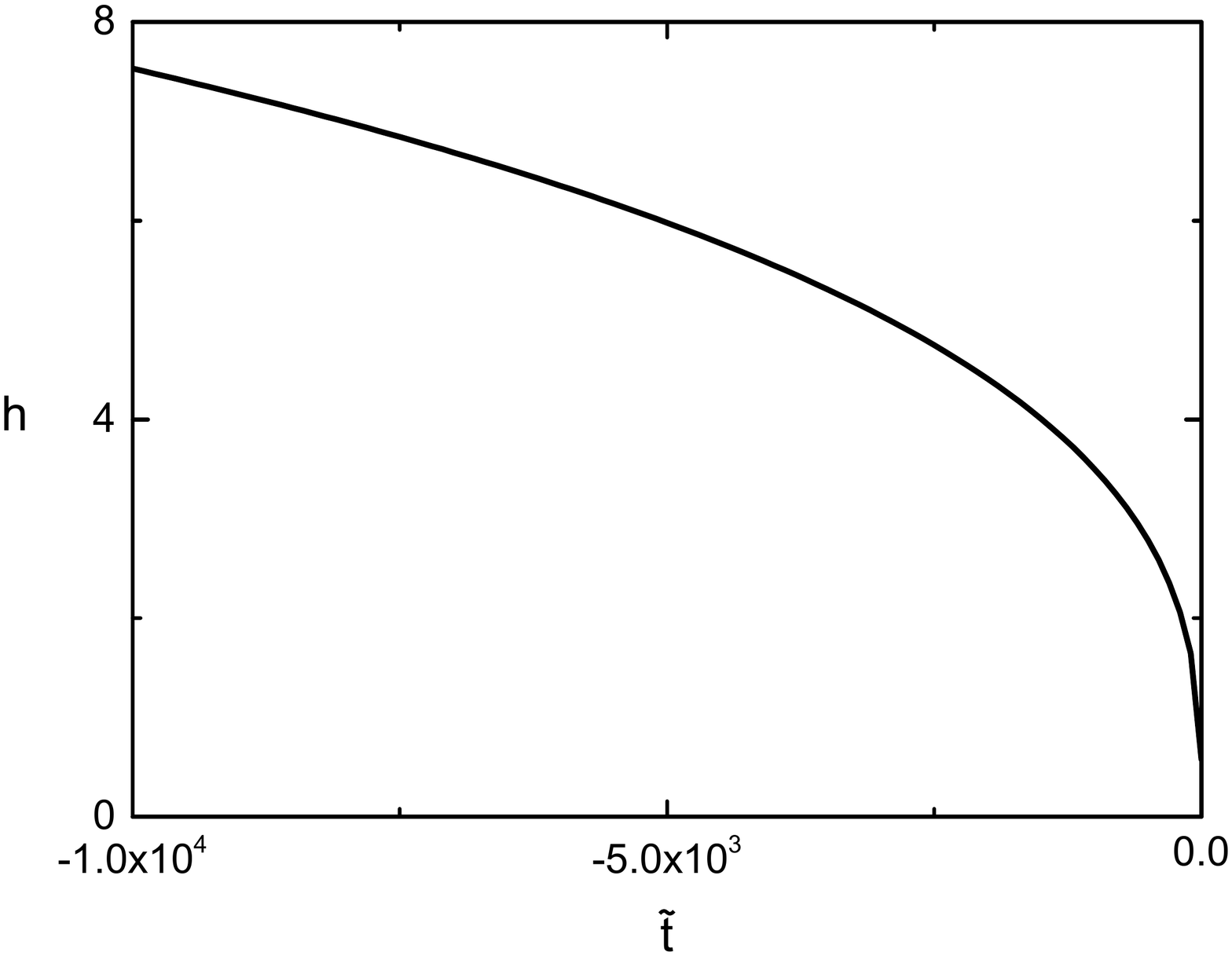}
\hfill
\includegraphics[width=.45\textwidth]{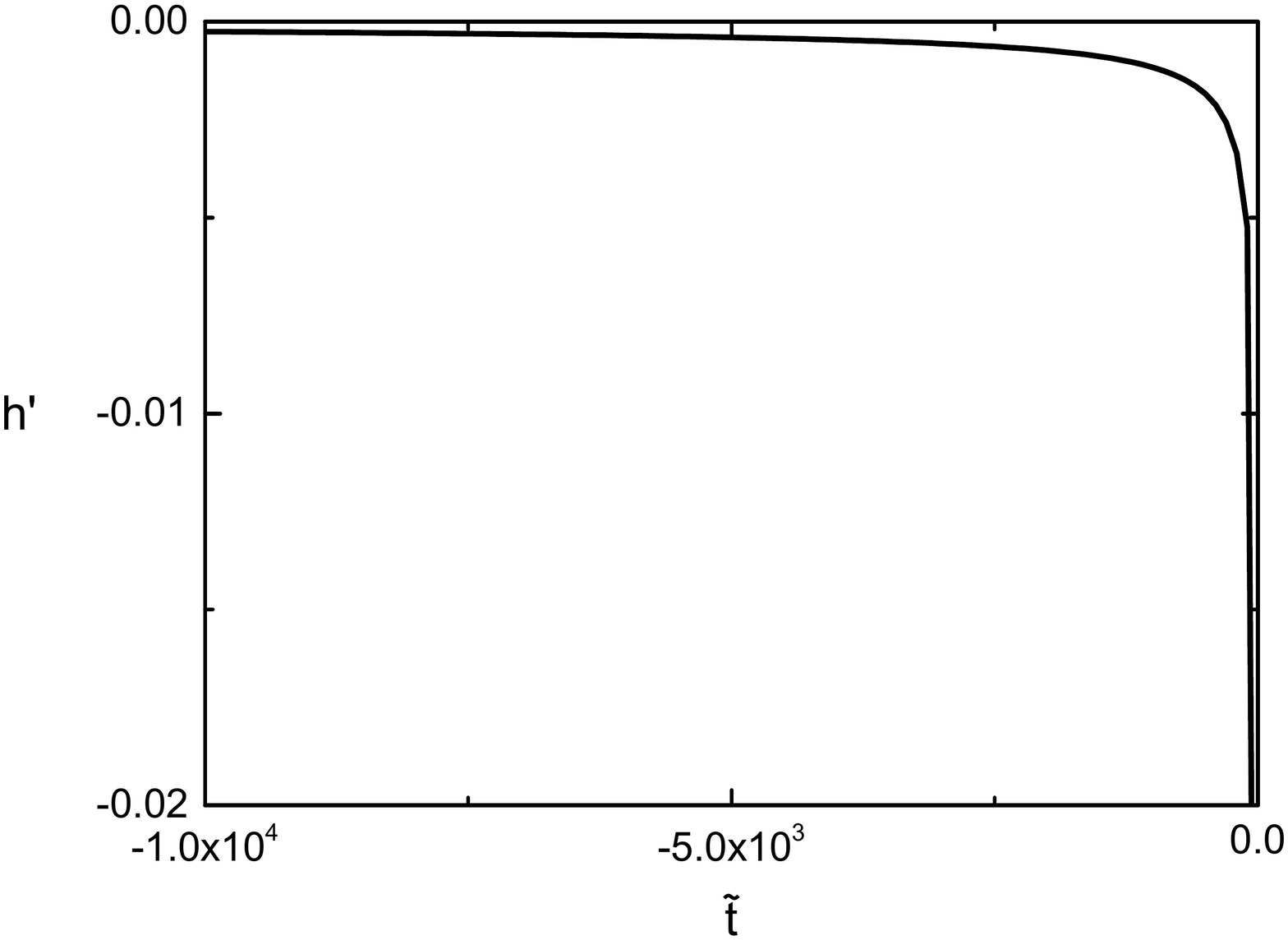}
% "\includegraphics" is very powerful; the graphicx package is already loaded
\caption{\label{fig:i} The evolution of Higgs field $h$ and its time derivative with respect to $\tilde{t}$
 in the contracting universe. We choose the parameters as in (\ref{parameters}), and the
 initial conditions of $h$ and $h'$ are $h_i=7.5321 M_p$, $h'_i=-2.5096\times10^{-4} M_p^2$.}
\label{higgsconplot}
\end{figure}

\begin{figure}[tbp]
\centering % \begin{center}/\end{center} takes some additional vertical space
\includegraphics[width=.45\textwidth]{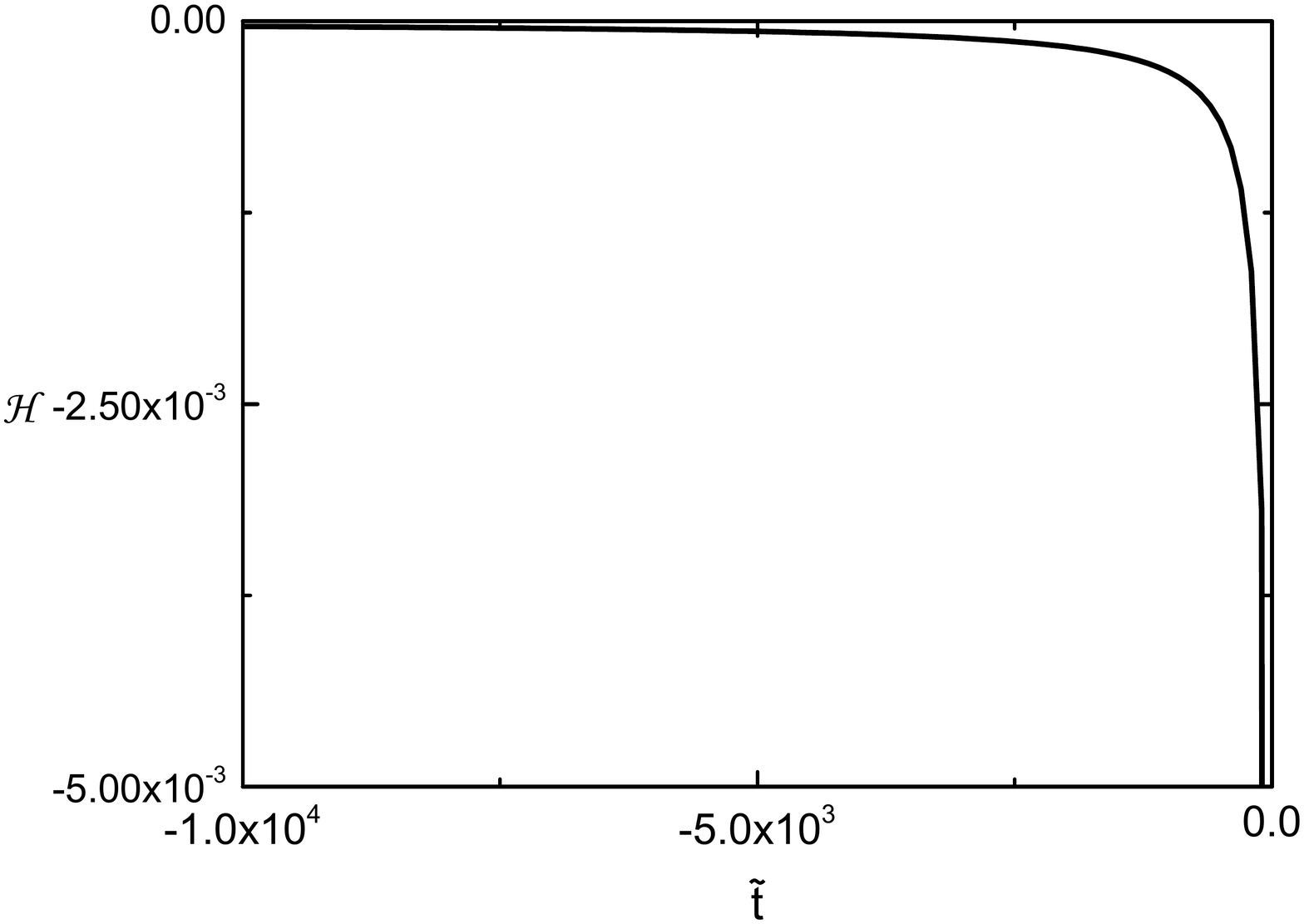}
\hfill
\includegraphics[width=.45\textwidth]{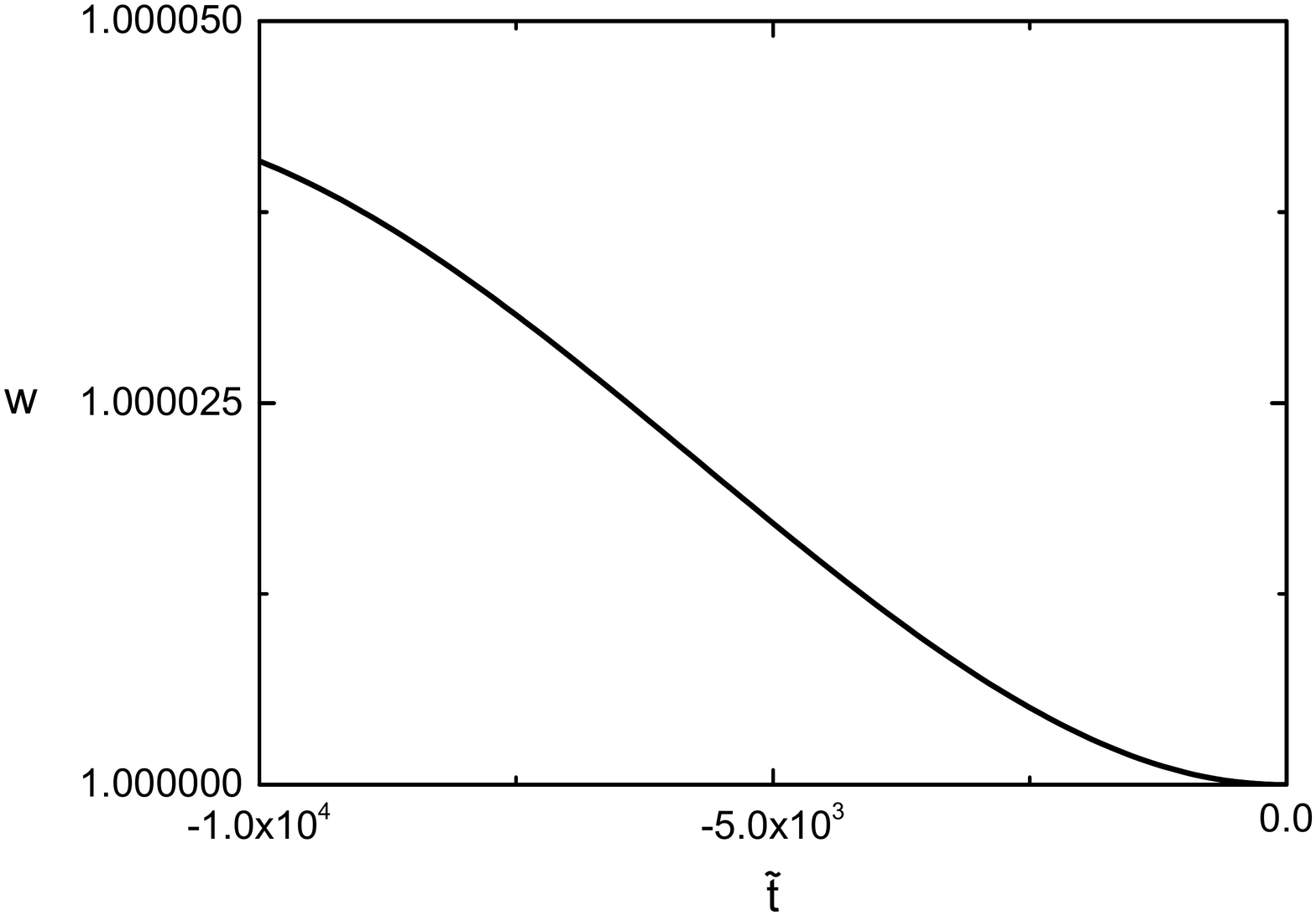}
% "\includegraphics" is very powerful; the graphicx package is already loaded
\caption{\label{fig:i} The evolution of the Hubble parameter ${\cal H}$ and Higgs EoS parameter $w$ with
respect to $\tilde{t}$ in the contracting universe. The same parameters and initial conditions
have been chosen as in Figure \ref{higgsconplot}.}
\label{hubbleconplot}
\end{figure}

It is difficult to solve the nonlinear differential equation (\ref{eomcon2}) analytically,
so we conduct numerical calculation. We solve the cosmological equations (\ref{CosmologyEqs}),
and show the evolutions of the Higgs field $h$, the Hubble parameter ${\cal H}$ as well
as the EoS parameter $w$ in contracting phase in Figures \ref{higgsconplot} and \ref{hubbleconplot}.
In our numerical calculation, one can see that we are starting with a contracting phase with a negative
Hubble parameter. The value of $h$ decreases, which means the Higgs field is moving leftwards along
the potential with a negative velocity. The EoS parameter is only slightly larger than unity,
which means that the kinetic energy is much larger than the potential energy, mainly
because of the large prefactor $3\xi^2h^2/M_p^2$ in front of the kinetic energy in (\ref{rhoprescon2}).
One can also see that the Hubble parameter becomes more and more negative. Although for the Higgs
field of the form (\ref{ActionEFRameConExp}), ${\cal H}'=-\frac{1}{2M_p^2}(\rho+p)$ is negative
definite where $\rho$ and $p$ are given in Eqs. (\ref{rhocon}) and (\ref{prescon}), as can be seen
in the next paragraph, the involvement of the Galileon term will make ${\cal H}'>0$, so that the
bounce can be triggered. Moreover, the final values of $h$ and $h'$ at the end of the contracting
phase can also be evaluated, which are:
\begin{equation}
\label{hminus}
h_-=0.5694 M_p~,~~~h_-'=-0.04391 M_p^2~.
\end{equation}

We'd like to make some discussion about the initial conditions of $h_i$ and $h'_i$. They are
fine-tuned to get a nonsingular bounce and a slow-roll inflation.  We argue that
$h'_i$ shouldn't be too small, or the Higgs field cannot roll into the false vacuum around
$h=0$, but will fall into the true vacuum around $h=0.75 M_p$. Also it is shouldn't be
too large, since a large velocity will not lead to slow-roll inflation after the
cosmological bounce.

\subsection{Bouncing phase}
It has been shown in the paragraph above that an anisotropy-free contraction can be realized
with the Higgs field action (\ref{ActionEFRameConExp}). However, in order to trigger the bounce
after the contraction, extra terms such as the Galileon term in (\ref{ActionEFrame}) have to be
applied.  Since the Galileon term contains acceleration of the field, thus in usual G-bounce
models, once the G-term dominates the whole Lagrangian, it will grow larger and larger, and it
will be very difficult to force the Galileon term to sub-dominant again, thus would lead to a
faster and faster expansion of the universe and a big rip singularity. This problem can be evaded
if we insert a delta/peak-like factor in front of the G-term in order to suppress after bounce,
this idea has been used in earlier works\cite{Cai:2012va,Koehn:2013upa,Qiu:2015nha}. In this paper
we consider the factor $c(h)$ to be as Eq. (\ref{functions}). Therefore, near the bounce point, the
Lagrangian can be written as
\begin{eqnarray}
\label{actionEFramebounce}
S_E&=&\int d\tilde{x}^4 \sqrt{-\tilde{g}}
\left\{-\frac{M_p^2}{2}\tilde{R}+\frac{6\xi^2}{M_p^2}h^2\tilde{X}A^{-2}(h)+\tilde{X}A^{-1}(h)-V(h)A^{-2}(h)\right. \nonumber\\
&&\left.+\frac{\alpha}{M_p^4}\left(\tilde{X}h\tilde{\Box}{h}-4\xi\frac{h^2}{M_p^2}\tilde{X}^2A^{-1}(h)\right)+\frac{\beta}{M_p^4} \tilde{X}^2\right\}~.
\end{eqnarray}
We write down the resulting energy and pressure as:
\begin{eqnarray}
\rho&=&\rho_{\rm k}+\rho_{\rm p}+\rho_{\rm e}~,\\
p&=&p_{\rm k}+p_{\rm p}+p_{\rm e}~,
\end{eqnarray}
where we have divided the energy density and pressure of the Higgs field into its normal kinetic
part, potential part and extra part (the second line of the action (\ref{actionEFramebounce})).
Therefore we have:
\begin{eqnarray}
\rho_{\rm k}&=&\left\{\frac{h'^2}{2}+\frac{3\xi^2}{M_p^2}h^2h'^2A^{-1}(h)\right\}A^{-1}(h)~,\\
p_{\rm k}&=&\left\{\frac{h'^2}{2}+\frac{3\xi^2}{M_p^2}h^2h'^2A^{-1}(h)\right\}A^{-1}(h)~,
\end{eqnarray}

\begin{eqnarray}
\rho_{\rm p}&=&V(h)A^{-2}(h),\\
p_{\rm p}&=&-V(h)A^{-2}(h),
\end{eqnarray}

\begin{eqnarray}
\rho_{\rm e}&=&-\frac{3\alpha\xi}{M_p^6}h^2h'^4A^{-1}(h)
+\left\{3{\cal H}h'^3h-\frac{h'^4}{2}\right\}\frac{\alpha}{M_p^4}
+\frac{3 \beta}{4 M_p^4}h'^{4}~,\\
p_{\rm e}&=&-\frac{\alpha\xi}{M_p^6}h^2h'^4A^{-1}(h)
+\left\{-h h'^2 h''-\frac{h'^4}{2}\right\}\frac{\alpha}{M_p^4}+\frac{\beta}{4 M_p^4}h'^{4}~.
\end{eqnarray}

We have numerically solved Eq.(\ref{CosmologyEqs}) again, and find that a nonsingular bounce does
happen. At the bounce point, the total energy is zero, which we have found in our numerical
calculation to be because of the cancelation between $\rho_{\rm e}$ and $\rho_{\rm k}$, which
dominate the total energy density in this phase, while the potential contribute very little to the
energy. We present the relationship of $\rho_{\rm k}$, $\rho_{\rm p}$ and $\rho_{\rm e}$ in
Figure \ref{EnergyBounceplot}. We can see that the energy is canceled between $\rho_{\rm k}$ and
$\rho_{\rm e}$, while the potential is rather small and can be safely omitted. This means after
the bounce the Higgs field will be fast rolling, leading to a decrease of the power spectrum on
large scale.

\begin{figure}[tbp]
\centering % \begin{center}/\end{center} takes some additional vertical space
\includegraphics[width=.45\textwidth]{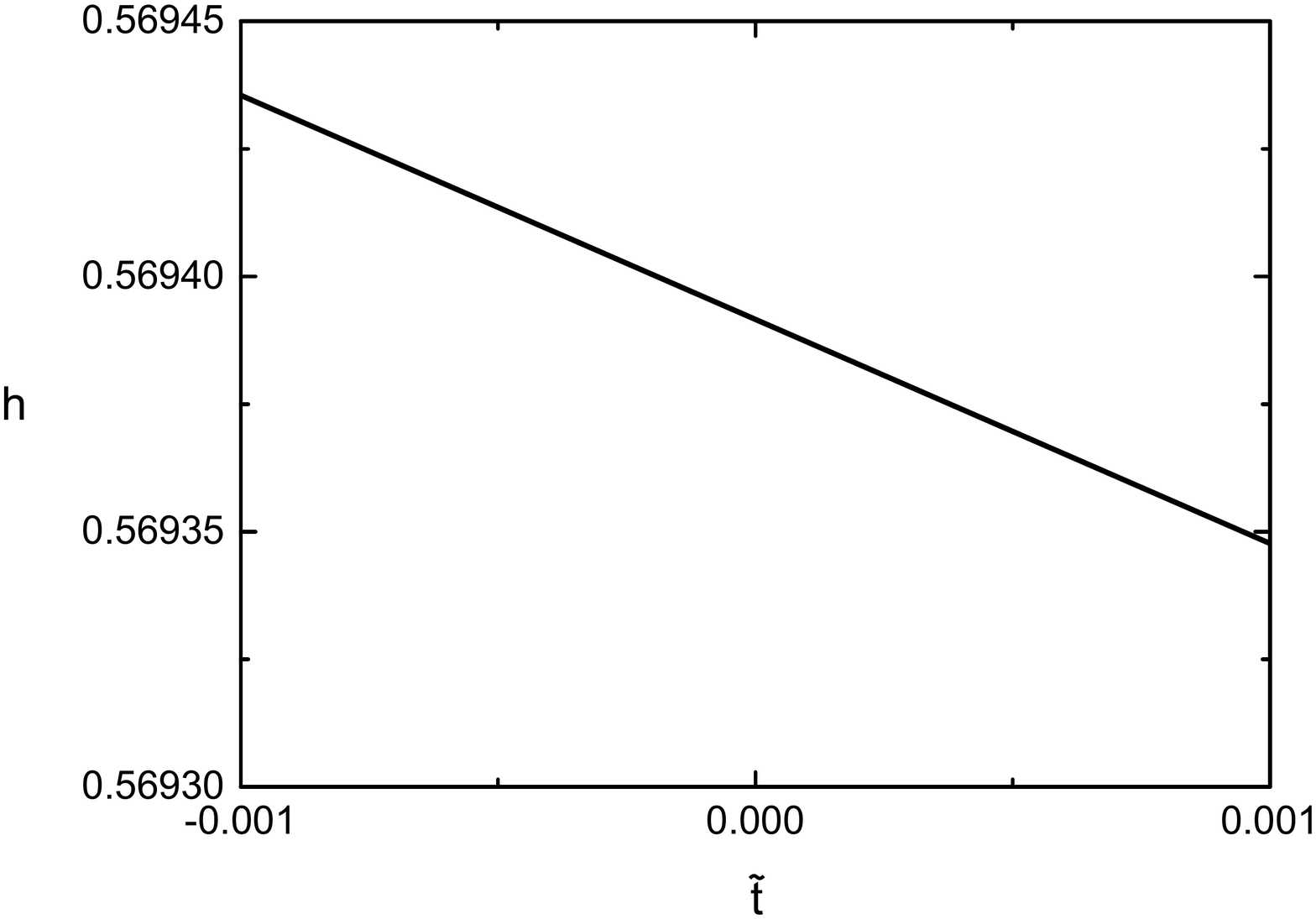}
\hfill
\includegraphics[width=.45\textwidth]{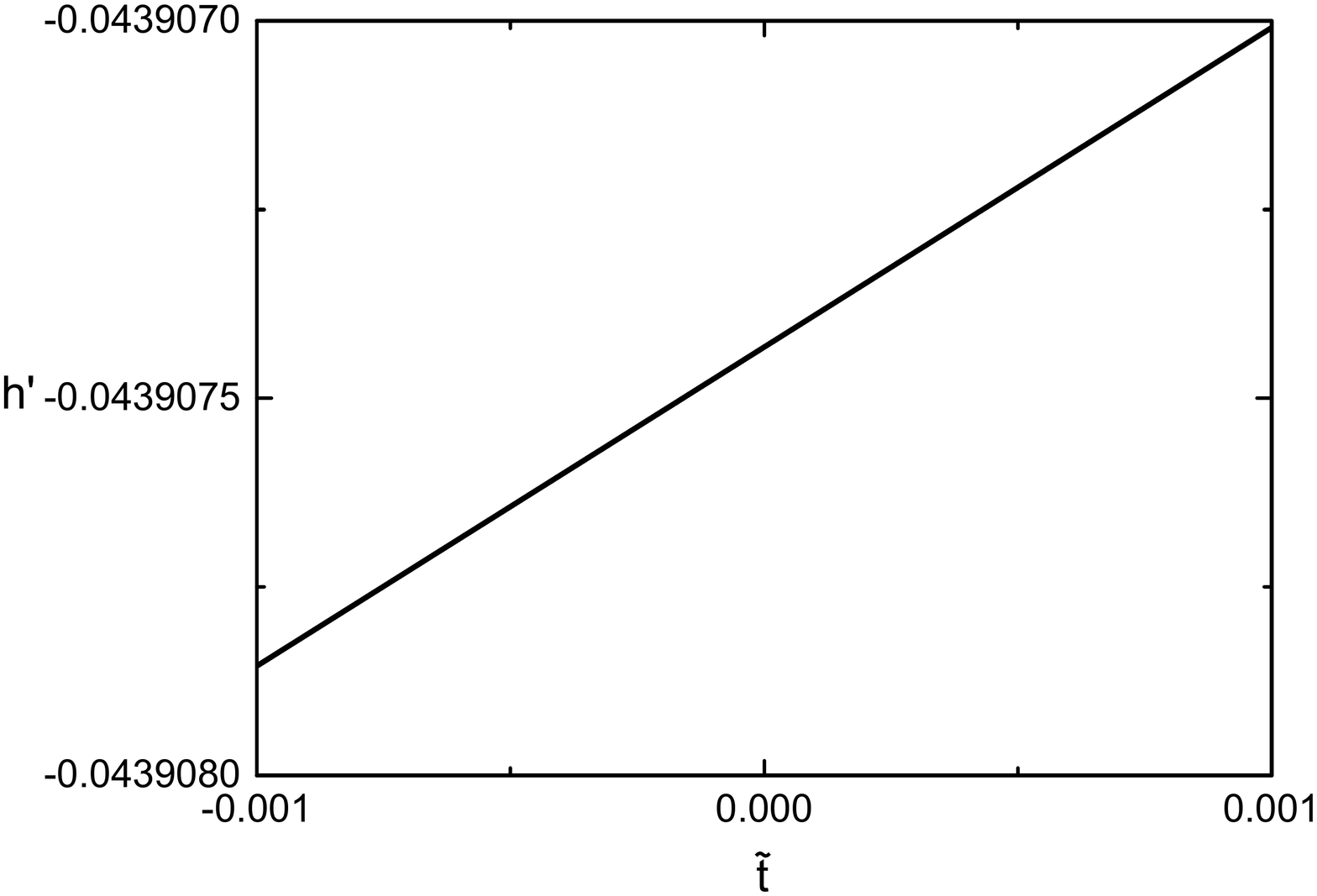}
% "\includegraphics" is very powerful; the graphicx package is already loaded
\caption{\label{fig:i} The evolution of Higgs field $h$ and its time derivative with respect to $\tilde{t}$
during the bounce. We choose the parameters as in (\ref{parameters}), and the initial conditions
of $h$ and $h'$ are $h_i=h_-$, $h_i'=h_-'$.}
\label{HiggsBounceplot}
\end{figure}

\begin{figure}[tbp]
\centering % \begin{center}/\end{center} takes some additional vertical space
\includegraphics[width=.45\textwidth]{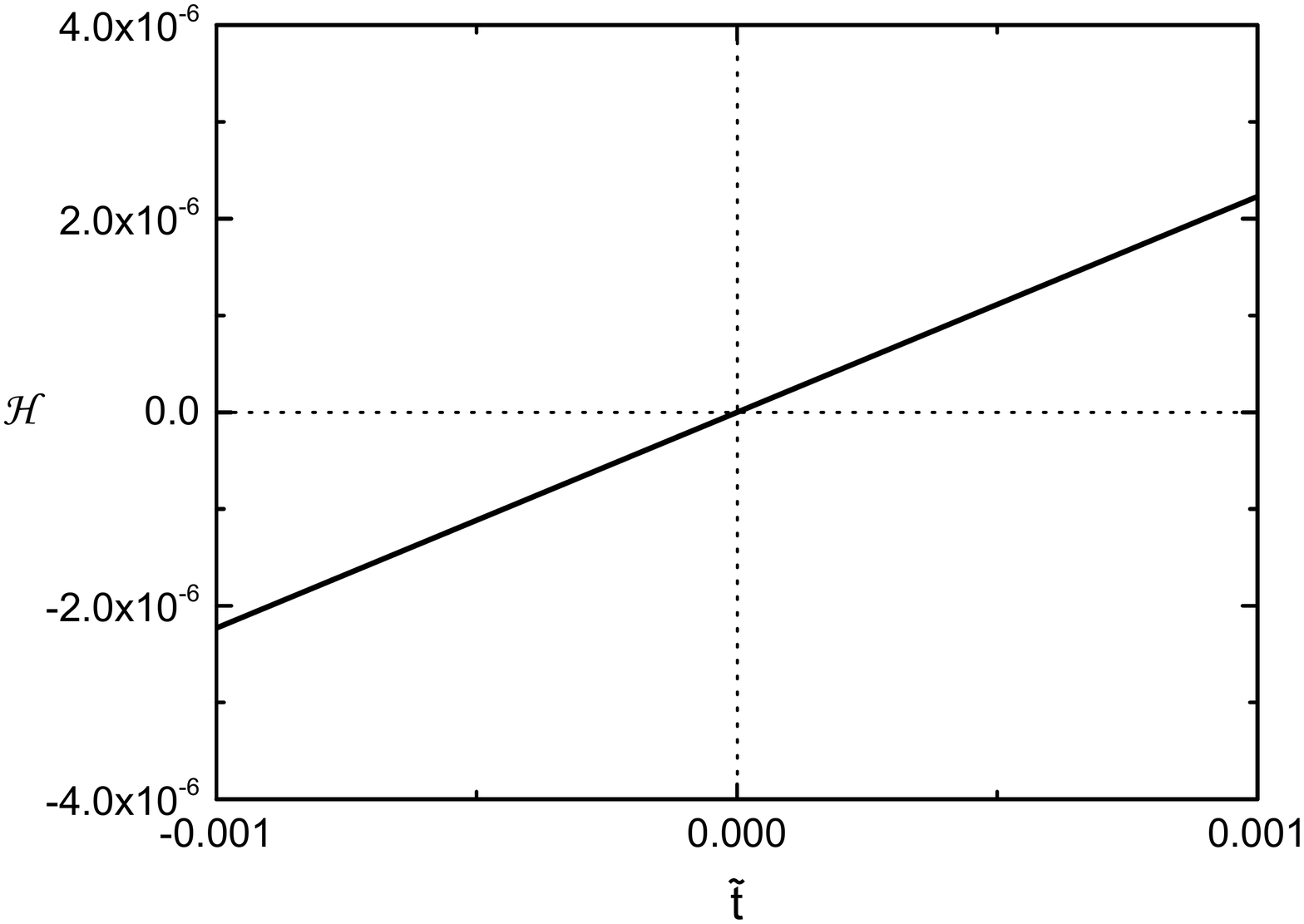}
\hfill
\includegraphics[width=.45\textwidth]{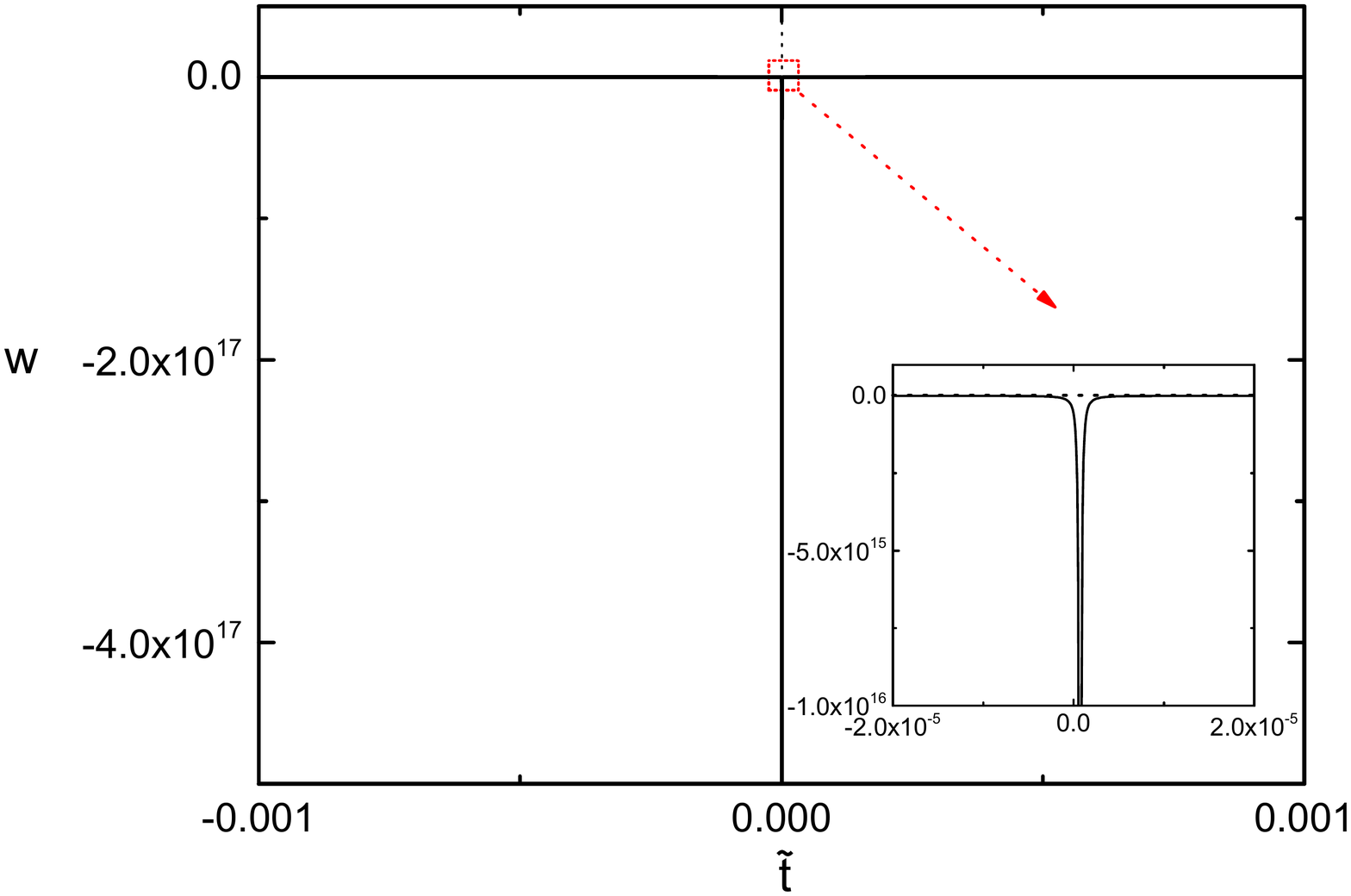}
% "\includegraphics" is very powerful; the graphicx package is already loaded
\caption{\label{fig:i} (Color plot.) The evolution of the Hubble parameter ${\cal H}$ and the Higgs EoS 
parameter $w$ with respect to $\tilde{t}$ during the bounce. The same parameters and initial conditions have
been chosen as in Figure \ref{HiggsBounceplot}. The EoS parameter around the bounce point has been zoomed in.}
\label{HubbleBounceplot}
\end{figure}

\begin{figure}[tbp]
\centering % \begin{center}/\end{center} takes some additional vertical space
\includegraphics[width=.65\textwidth]{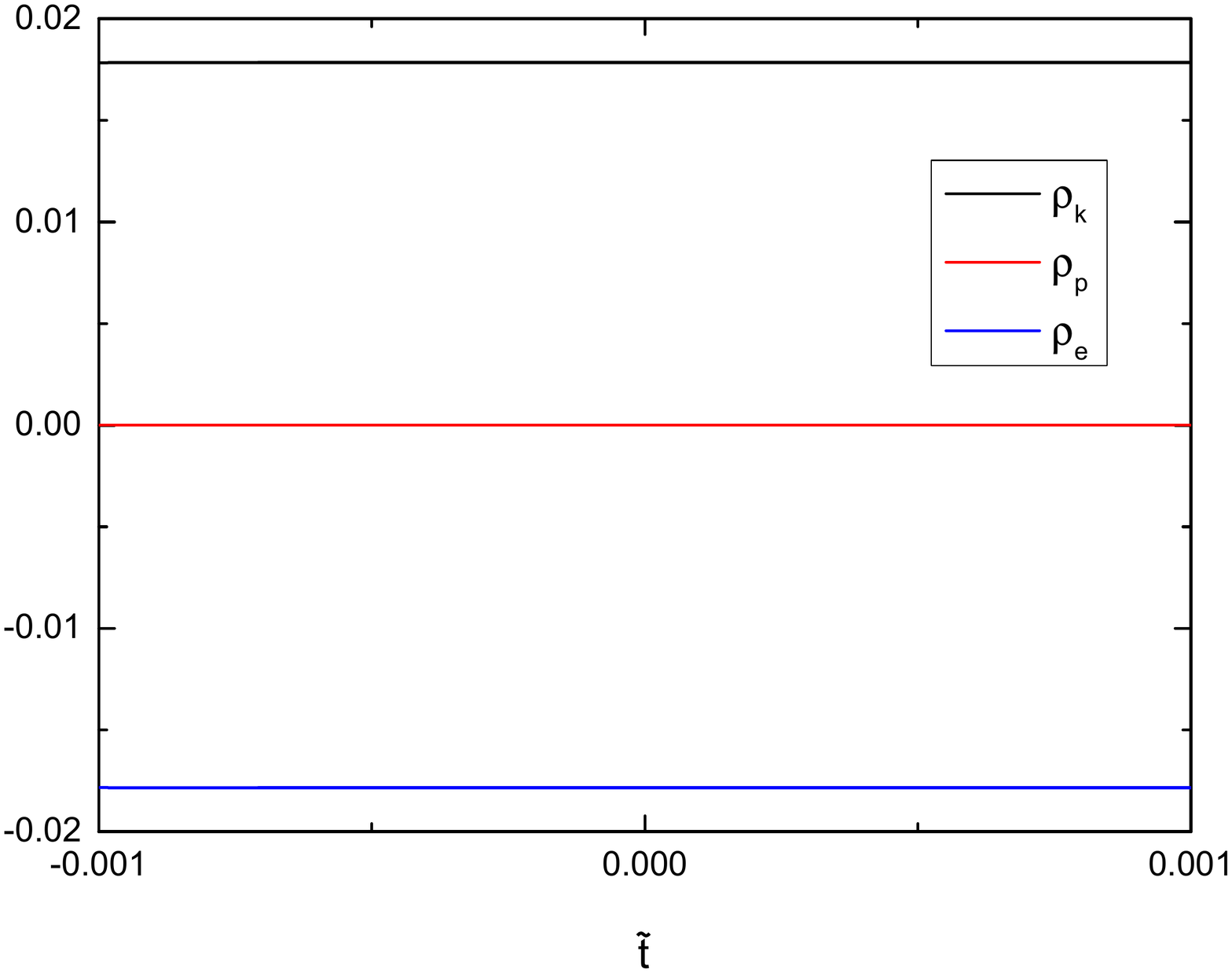}
% "\includegraphics" is very powerful; the graphicx package is already loaded
\caption{\label{fig:i} (Color plot.) Energy components in the contracting universe. The same parameters
and initial conditions have been chosen as in Figure \ref{HiggsBounceplot}. We can see that the
total energy is dominated and canceled by $\rho_k$ and $\rho_e$, while $\rho_p$ is
negligible.}
\label{EnergyBounceplot}
\end{figure}

In Figures \ref{HiggsBounceplot} and \ref{HubbleBounceplot}, we plot the evolution of the Higgs
field $h$ and its time derivative $h'$, the Hubble parameter ${\cal H}$ as well as the EoS parameter
$w$ in the bounce region. In order to keep the continuity of the evolution, in this region we
choose the initial condition of $h$ and $h'$ to be their final values in the contraction phase,
namely Eq.(\ref{hminus}). From the plots we can see that the Hubble parameter can transit almost
linearly from a negative value to a positive value, indicating that a bounce occur. Moreover,
the EoS parameter of the Higgs field is much smaller than $-1$, which means a violation of the
NEC. Furthermore we can see that, after the bounce, the Higgs field has a negative velocity,
so it will climb the potential hill towards the false vacuum. As it is climbing, its velocity
will be decrease for two reasons: the first is that some of its kinetic energy density is
transferred to potential energy, the second decrease is due to the friction effect of a positive
Hubble term in the equation of motion. This velocity decrease will make the kinetic energy of
Higgs fall below the potential energy, driving the Higgs field to experience a slow-roll
inflation after the bounce. The final values of $h$ and $h'$ at the end of the bouncing phase
can also be evaluated, and they are:
\begin{equation}
\label{hplus}
h_+=0.56934768 M_p~,~~~h_+'=-0.04390701 M_p^2~.
\end{equation}

\subsection{Expanding phase}

After the bounce, the universe goes to an expanding phase. Due to the delta function $c(h)$,
the G-term disappears again, so the Lagrangian is just the same one as shown in
Eq.(\ref{ActionEFRameConExp}). In this period, we still use the continuity condition of the Higgs
field and its derivative to determine the initial conditions, namely the evolution starts with
$h$ and $h'$ to be as in Eq.(\ref{hplus}). Now the Higgs has climbed over its maximum, it falls
down to the false vacuum in the center, only the first term in $V(h)$ dominates, so $V(h)$ becomes
the same as the normal Higgs potential without corrections. In the expanding phase with such a
potential, the evolution will be attracted to a slow-rolling one with tiny $h'$. The equations
for $\rho$, $p$ and the EoM should be the same as Eq.s (\ref{rhocon}), (\ref{prescon}) and (\ref{eomcon}),
however, different from the contracting phase, the slow-roll condition can be applied which greatly
simplifies our calculations. The slow-roll conditions yield:
\begin{equation}
3{\cal H}h'\simeq-\frac{\partial(A^{-2}(h)V(h))}{\partial h}\simeq \frac{\lambda M_p^6 h^3}{(M_p^2+\xi h^2)^3}~,~~~3M_p^2{\cal H}^2\simeq A^{-2}(h)V(h)\simeq \frac{\lambda M_p^4 h^4}{4(M_p^2+\xi h^2)^2}~
\end{equation}
where by using the approximation $\xi h^2/M_p^2\gg1$ and the specific value in (\ref{parameters}),
one can determine the Hubble parameter during inflation:
\begin{equation}
\label{hubblevalue}
{\cal H}_{inf}\simeq\sqrt{\frac{\lambda}{12}}\frac{M_p}{\xi}\simeq 3.456\times 10^{-7} M_p~.
\end{equation}

To do slow-roll analysis, we'd like to redefine a new scalar field $\chi$ to get
simpler equations
\begin{eqnarray}
\frac{d\chi}{dh}\equiv \sqrt{\frac{\Omega^2+6\xi^2 h^2/M_p^2}{\Omega^4}},
\end{eqnarray}
then the potential reads
\begin{eqnarray}
U\left(\chi\right) = V\left(h(\chi)\right)A^{-2}\left(h(\chi)\right).
\end{eqnarray}

Defining the slow-roll parameters
\begin{equation}
\epsilon = \frac{M_p^2}{2}\left(\frac{U_{,\chi}}{U}\right)^2~,~~~\eta = M_p^2 \frac{U_{,\chi\chi}}{U}~,
\end{equation}
and using the condition $\epsilon_f=1$, one can solve for the final value of $h$:
\begin{equation}
\label{hfinal}
h_f=0.0006 M_p,
\end{equation}
it deviates a little bit from the numerical solution $h_f=0.0014 M_p$ .
Moreover, the e-folding number $N$ can be calculated through:
\begin{equation}
\label{Nvalue}
N = \frac{1}{M_p^2}\int^{\chi_{\rm i}}_{\chi_{\rm f}}\frac{U}{U_{,\chi}}d\chi.
\end{equation}
Therefore using our initial condition (\ref{hplus}) as well as the final value (\ref{hfinal}),
one can evaluate the e-folding number of the inflation phase in our model, $N=59.4$.

We can also briefly estimate the perturbations that could be obtained in our model. If we neglect
the effects from bounce and contraction phase, i.e., we focus on the perturbations which are only
generated in the inflationary phase, thus, from \cite{Mukhanov:1990me}, the scalar and tensor power
spectra can be found as:
\begin{equation}
P_{\zeta} = \frac{U\left(\chi\right)}{24\pi^2 M_p^4 \epsilon}~,~~~P_{T} = \frac{2 U\left(\chi\right)}{3\pi^2 M_p^4}~,
\end{equation}
and the two very important observables---the index of power spectrum and the tensor-scalar ratio
are:
\begin{equation}
n_s \equiv 1+\frac{d\ln{P_{\zeta}}}{d\ln{k}}\sim 1-6\epsilon+2\eta~,~~~r \equiv \frac{P_{\zeta}}{P_{ T}}\sim16\epsilon
\end{equation}
which are the standard results. Of course one should also take into account the effects
from the bounce or contraction phase. However, as has been demonstrated in \cite{Piao:2003zm,Qiu:2015nha},
for bounce inflation model with $w>1$ in contracting phase, even some of the large scale fluctuation
modes will be generated before bounce, they will not exit the horizon. Therefore it will only bring
some mild corrections on the spectrum, such as some oscillating behavior, but will not alter the result
too much. For fluctuations on very large scales that exit the horizon before the bounce, blue-tilted
primordial spectrum will be obtained, which can suppress the CMB TT spectrum, but most of them can
only be observed by future observations. The detailed analysis of the perturbations of our model
including the pre-inflationary effects will be performed in our coming work.

\begin{figure}[tbp]
\centering % \begin{center}/\end{center} takes some additional vertical space
\includegraphics[width=.45\textwidth]{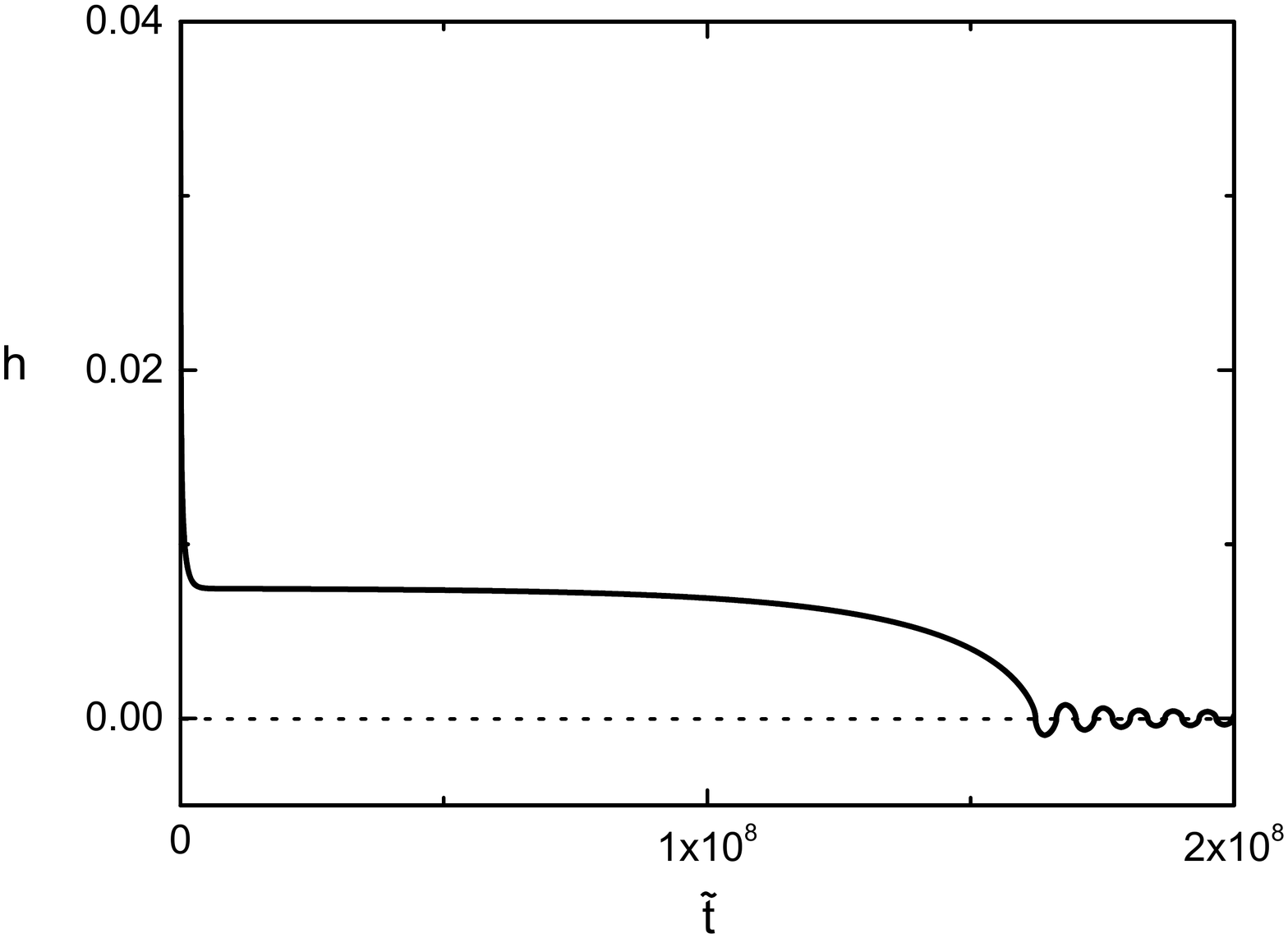}
\hfill
\includegraphics[width=.45\textwidth]{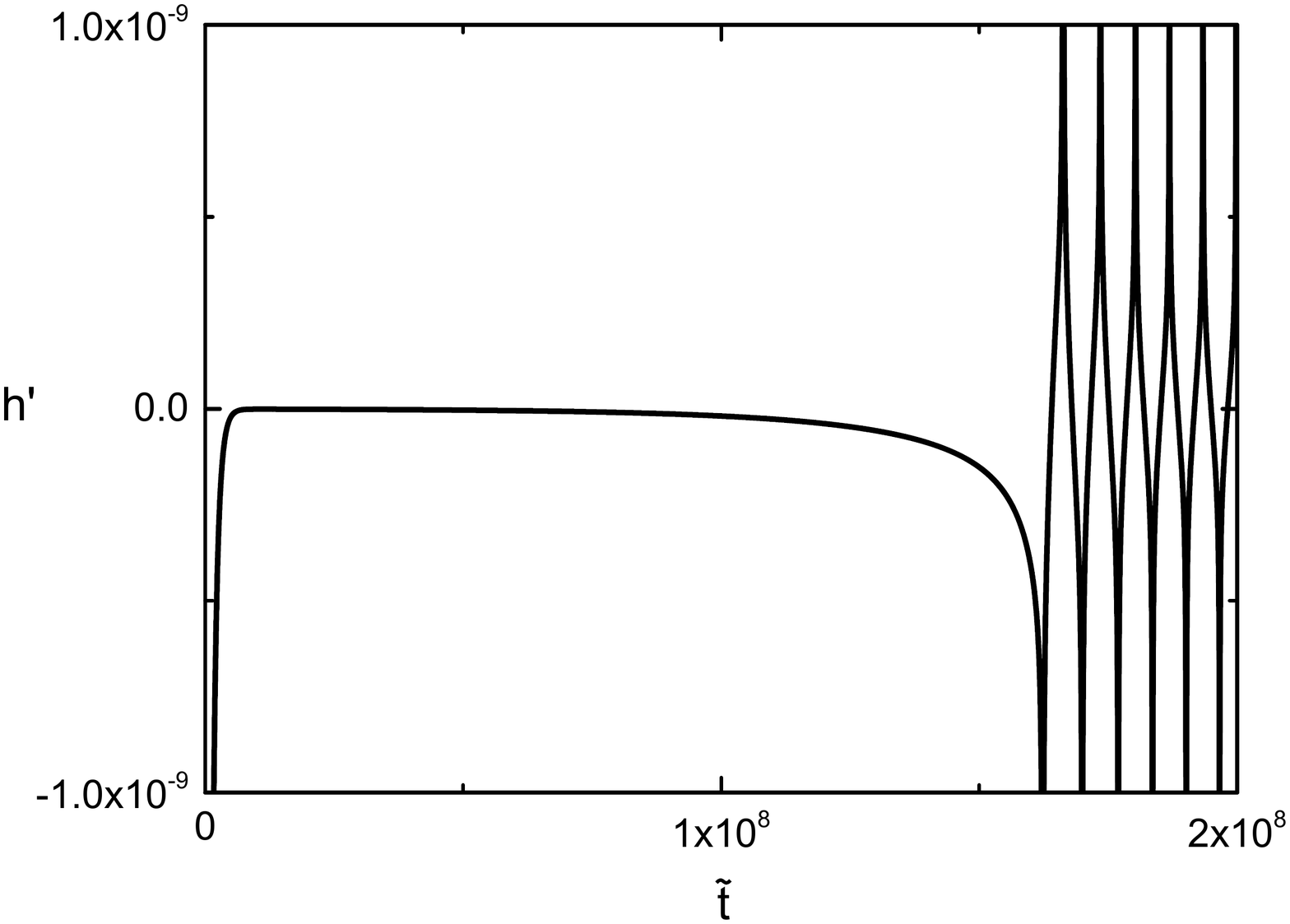}
% "\includegraphics" is very powerful; the graphicx package is already loaded
\caption{\label{fig:i} The evolution of Higgs field $h$ and its time derivative with respect to $\tilde{t}$
in expanding phase. We choose the parameters as in (\ref{parameters}), and the initial
conditions of $h$ and $h'$ are $h_i=h_+$, $h_i'=h_+'$.}
\label{higgsexpplot}
\end{figure}

\begin{figure}[tbp]
\centering % \begin{center}/\end{center} takes some additional vertical space
\includegraphics[width=.45\textwidth]{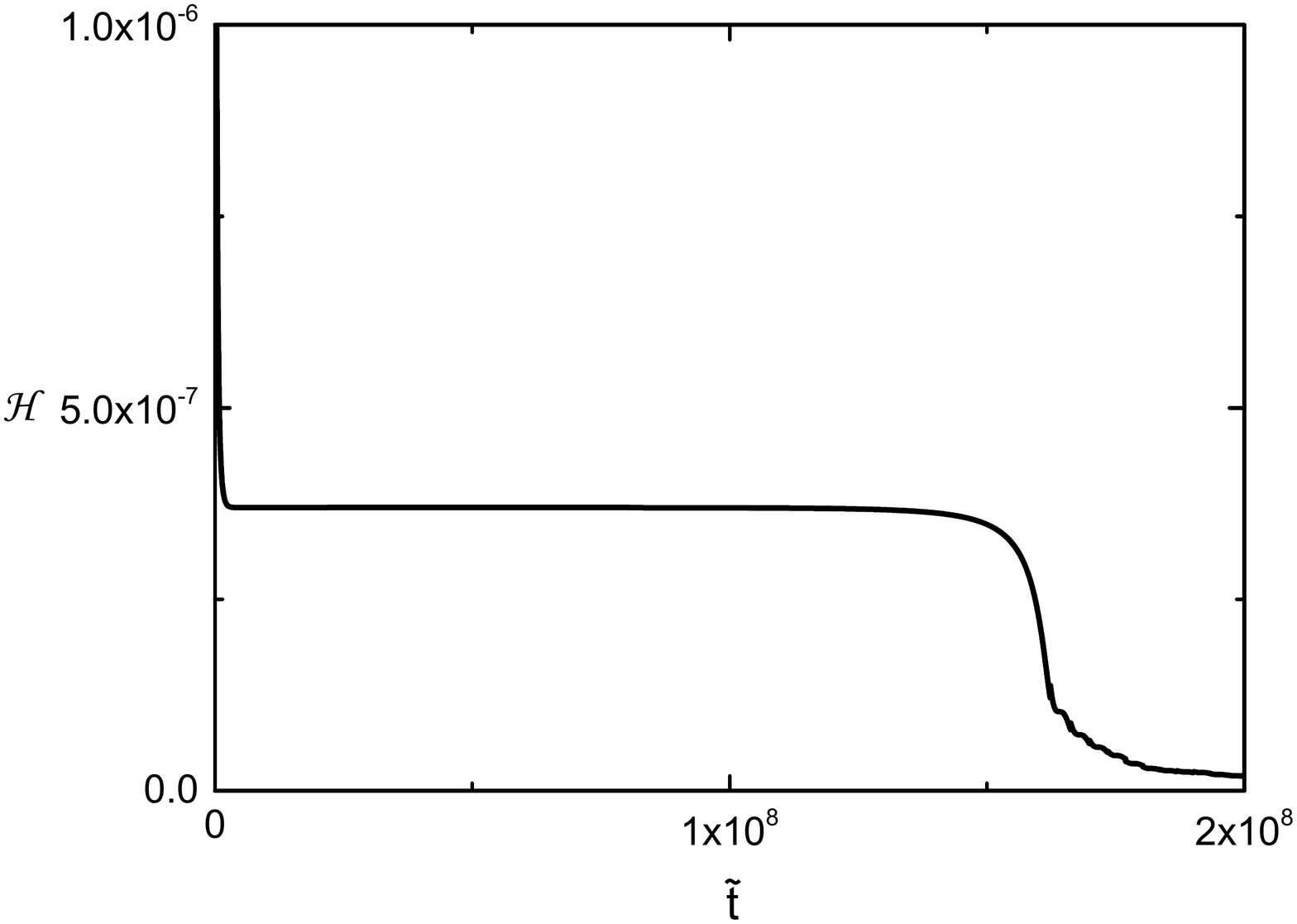}
\hfill
\includegraphics[width=.45\textwidth]{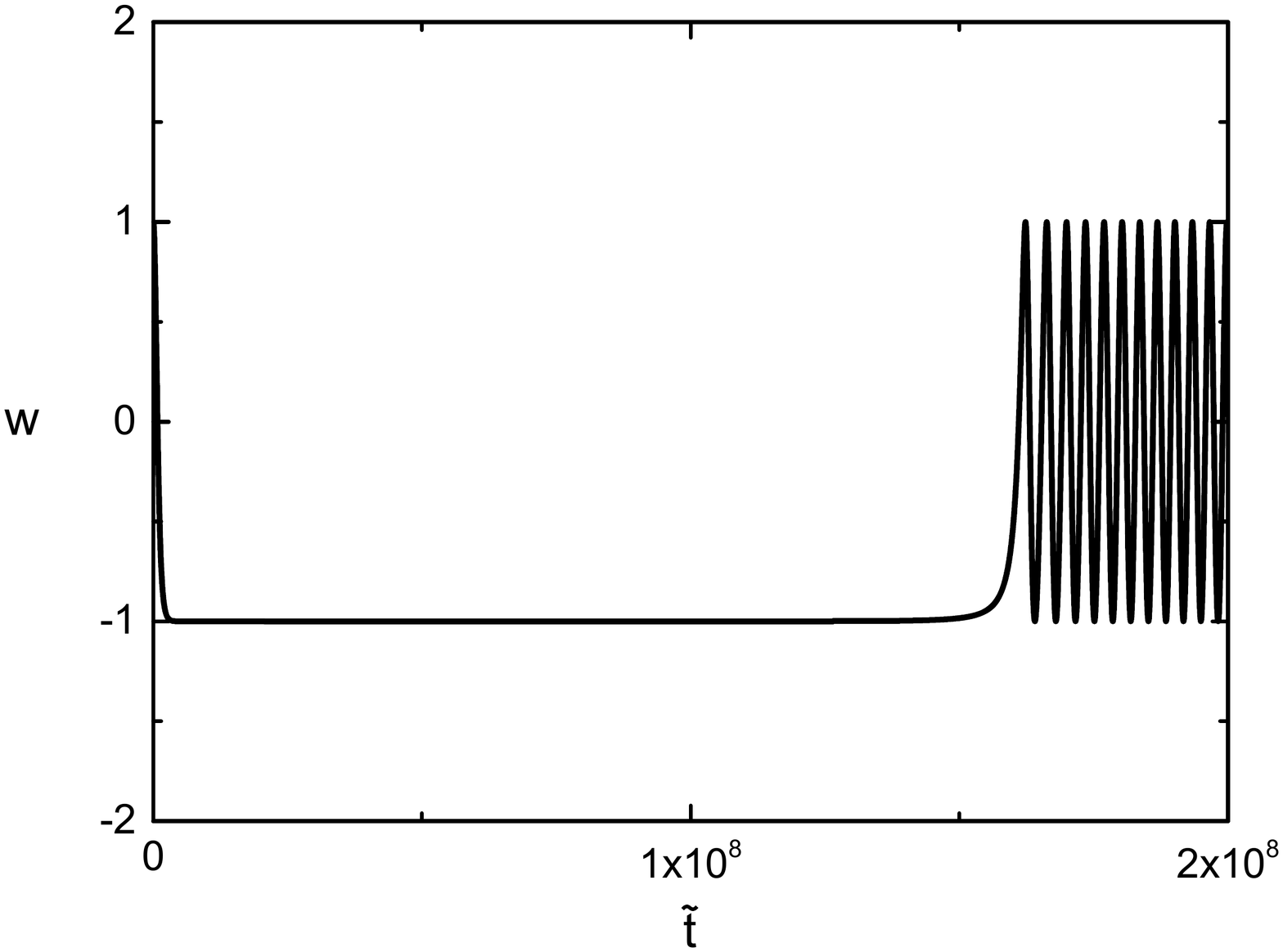}
% "\includegraphics" is very powerful; the graphicx package is already loaded
\caption{\label{fig:i} The evolution of the Hubble parameter ${\cal H}$ and Higgs EoS parameter $w$ with
respect to $\tilde{t}$ in expanding phase. The same parameters and initial conditions have
been chosen as in Figure \ref{higgsexpplot}.}
\label{hubbleexpplot}
\end{figure}

In Figures \ref{higgsexpplot} and \ref{hubbleexpplot} we numerically solve the evolution of $h$,
$h'$, ${\cal H}$ as well as $w$ in the expanding universe. We can see from the figures that after
the bounce, the Higgs field will soon be attracted into a position of $h\simeq 0.0074 M_p$,
$h'\simeq -2.33\times10^{-12} M_p^2$. Because of such a tiny velocity, the field slowly rolls down to
the false vacuum of the potential, driving inflation. During inflation the Hubble parameter is
nearly constant with the value about $3.69\times 10^{-7} M_p$, consistent with the analytical result
(\ref{hubblevalue}). Moreover, its EoS parameter $w$ is almost equal to $-1$.  After the time at
$1.6\times 10^{8}M_p^{-1}$, all the quantities begin to oscillate, which implies that the field has
reached the false vacuum and the universe enter the reheating phase. From the plot we can also
get that the e-folding number is $N\sim {\cal H}\cdot \Delta \tilde{t}\simeq 58$, which is
consistent with our analytical result (\ref{Nvalue}).

Our solution gives a tensor-scalar ratio $r=1.2\times10^{-5}$ and power index $n_s=0.84$.
In our model, there are only two parameters, the cut off $\Lambda$ and the non-minimal coupling
constant $\xi$, which are relevant to inflation. We have scanned the parameter space without
finding the right number of $n_s$ which is needed by observation. In fact, a inflation model
with $\lambda h^4$ plus a Coleman-Weinberg type potential can not generate the right number
of $n_s=0.96$ if $\Lambda$ is smaller than $M_p$.

However, we expect that the potential exists which yields a bounce inflation model and generates
the right $r$ and $n_s$. The conditions of cosmological bounce is $\rho_{\rm tot}=0$ and the
NEC is violated at the bounce point. In our model, they are satisfied since $\rho_k$ and a
negative $\rho_e$ dominate the total energy density(and $p_k$ and $p_e$ dominate the total
pressure) near the bounce point, and $\rho_k+\rho_e\simeq0$, $p_k+p_e< 0$. It turns out that
the Higgs potential is irrelevant to the existence of a bounce, since it contributes a negligible
amount to the energy and the pressure near the bounce point. On the other hand, the Higgs
potential is absolutely relevant to the existence of a successful Higgs inflation model.
So the conditions of inflation and bounce are independent with each other. According to
Planck's newest constraint on inflation\cite{Ade:2015lrj}, the non-minimal coupling Higgs
Inflation Model still survives. So our model could be an oversimplified type of Higgs Bounce
Inflation model, after we deal with the Higgs effective potential thoroughly, we'd get the
right number of $r$ and $n_s$. We won't do this in this paper, since we'd like to take the
virtue of analytic formula of the Coleman-Weinberg potential, and to show how we get the bounce
clearly without paying too many attentions on the details of the complete form of the Higgs
effective potential.

\begin{figure}[tbp]
\centering % \begin{center}/\end{center} takes some additional vertical space
\includegraphics[width=.65\textwidth]{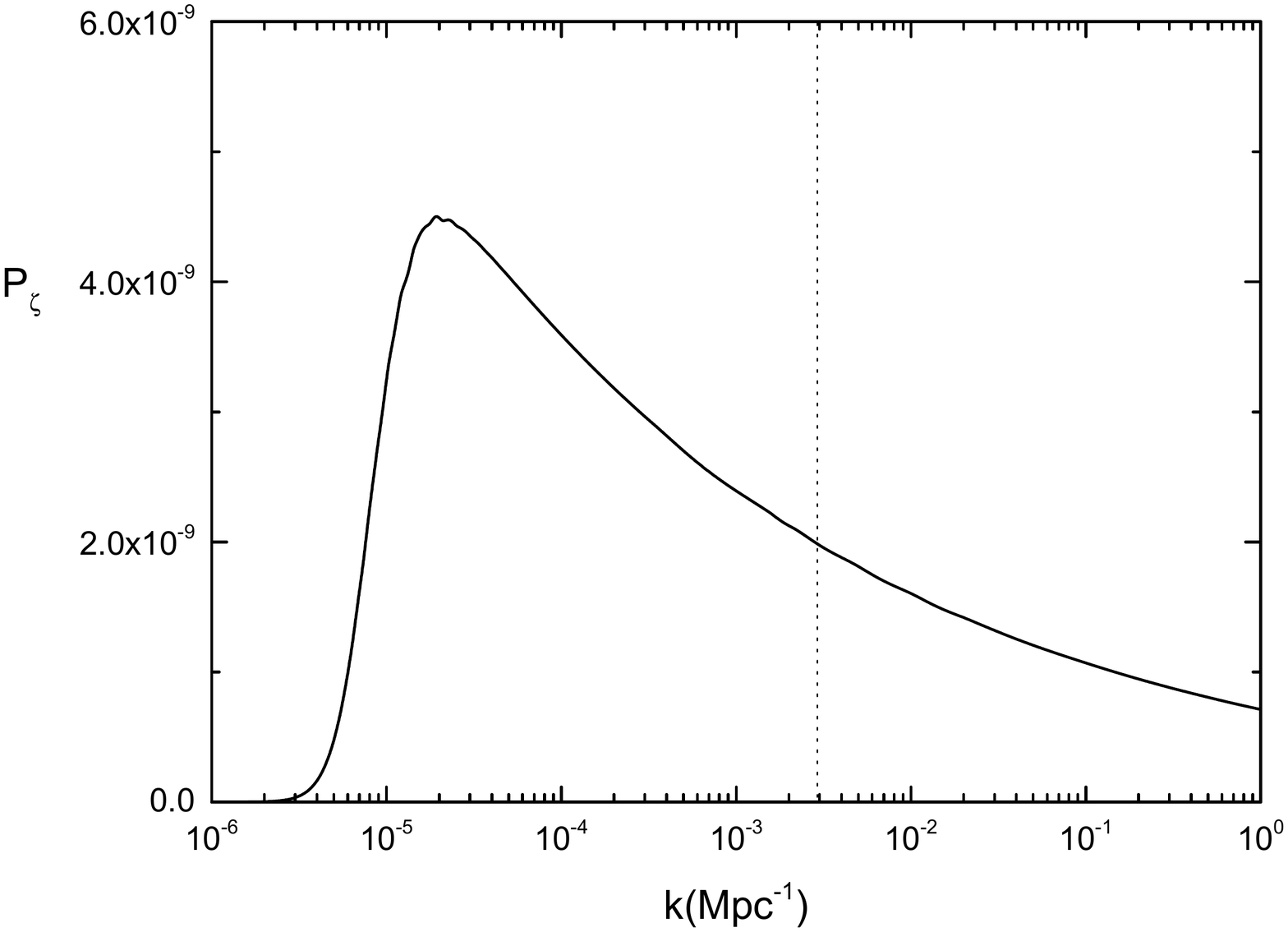}
% "\includegraphics" is very powerful; the graphicx package is already loaded
\caption{\label{fig:i} The power spectrum generated in expanding phase. The same parameters and initial
conditions have been chosen as in Figure \ref{higgsexpplot}. The dotted vertical line shows
the pivot scale, which is taken as $0.002 {\rm Mpc}^{-1}$\cite{Ade:2015lrj}.}
\label{psplot}
\end{figure}

In Figure \ref{psplot} we show the power spectrum generated during the expanding phase.
The vertical dotted line denotes the pivot scale, which has been taken as
$0.002 {\rm Mpc}^{-1}$ following Planck's paper\cite{Ade:2015lrj}. We can see that there
is a peak on large scale, we point that out it comes from the fast-roll period before
the slow-roll inflation. Notice this kind of fast-roll can explain the small $l$ anomaly
in CMB observation \cite{Hinshaw:2012aka, Contaldi:2003zv, Feng:2003zua, Liu:2010fm, Liu:2013kea, Xia:2014tda,Wan:2014fra}.

\section{Stability analysis}

Following \cite{Kobayashi:2010cm}, we can write down the quadratic action
\begin{equation}
\label{2nd-action}
S^{(2)}=\frac{1}{2}\int d\tilde{\tau}d\tilde{x}^3 z^2
\left[{\cal G}\left(\partial\zeta/\partial\tilde{\tau}\right)^2-
{\cal F} \left(\partial\zeta/\partial\tilde{x}\right)^2\right],
\end{equation}
in which $\zeta$ is the curvature perturbation in unitary gauge(in this gauge we have $\delta h =0$),
$c_s^2 \equiv \frac{{\cal F}}{{\cal G}}$ is the sound speed square, and
\begin{eqnarray}
z =
&&\frac{\tilde{a} h'}{{\cal H}-\frac{\alpha}{2 M_p^6}hh'^3},\\
\label{fff}
{\cal F} =
&&-\frac{2\alpha^2}{M_p^{10}}h^2 \tilde{X}^2
-\frac{8\alpha \xi}{M_p^{6}}h^2 A^{-1}(h)\tilde{X}
+\frac{2\beta}{M_p^{4}}\tilde{X}
+\frac{2\alpha}{M_p^{4}}h \left(h''+2{\cal H}h'\right)\\ \nonumber
&&+\frac{6\xi^2}{M_p^{2}}h^2 A^{-2}(h)+A^{-1}(h), \\
\label{ggg}
{\cal G} =
&&\frac{6\alpha^2}{M_p^{10}}h^2 \tilde{X}^2
-\frac{24\alpha \xi}{M_p^{6}}h^2 A^{-1}(h)\tilde{X}
+\frac{6\beta}{M_p^{4}}\tilde{X}
-\frac{4\alpha}{M_p^{4}}\tilde{X}
+\frac{6\alpha}{M_p^{4}}h {\cal H}h'\\ \nonumber
&&+\frac{6\xi^2}{M_p^{2}}h^2 A^{-2}(h) +A^{-1}(h).
\end{eqnarray}

\begin{figure}[tbp]
\centering % \begin{center}/\end{center} takes some additional vertical space
\includegraphics[width=.65\textwidth]{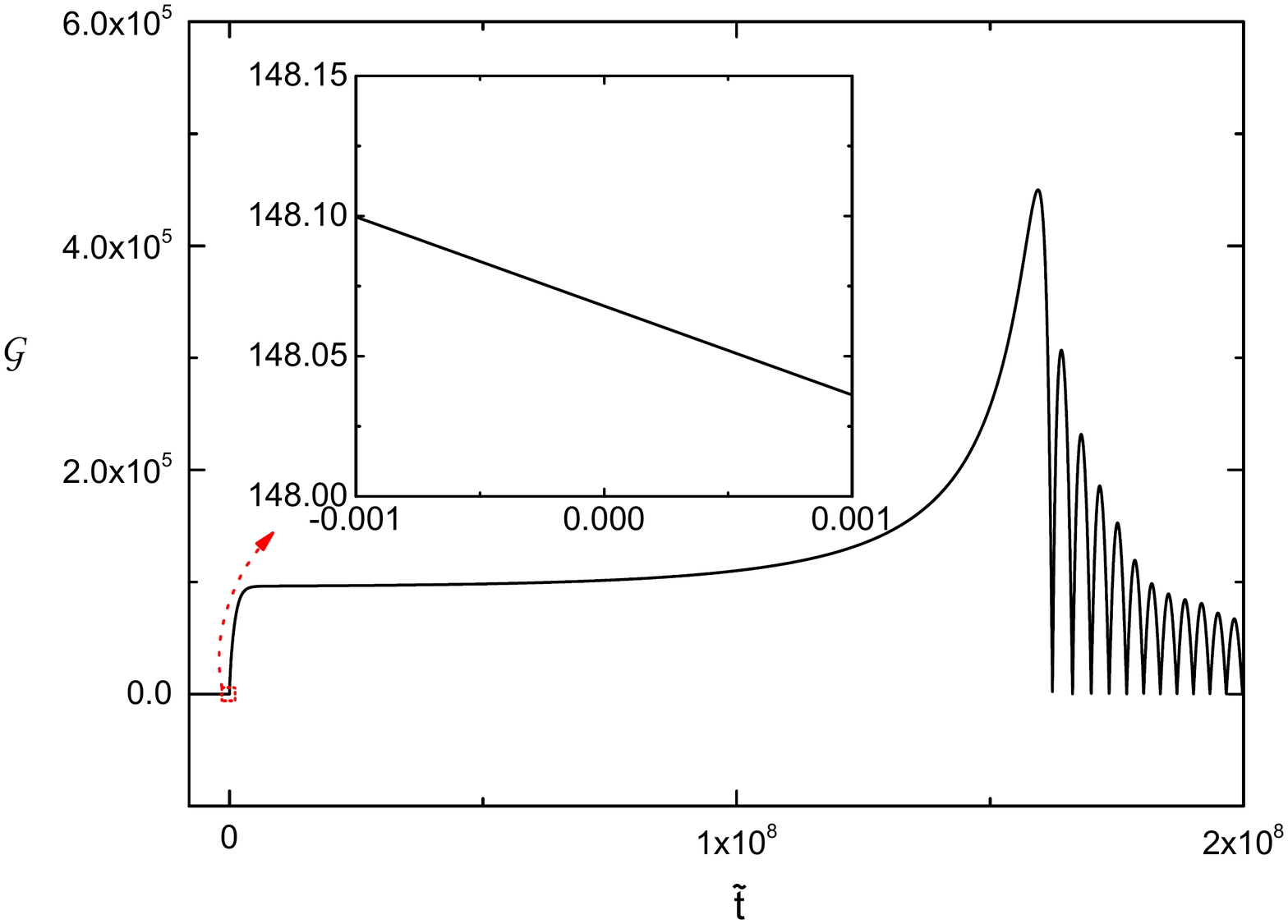}
\hfill
\includegraphics[width=.65\textwidth]{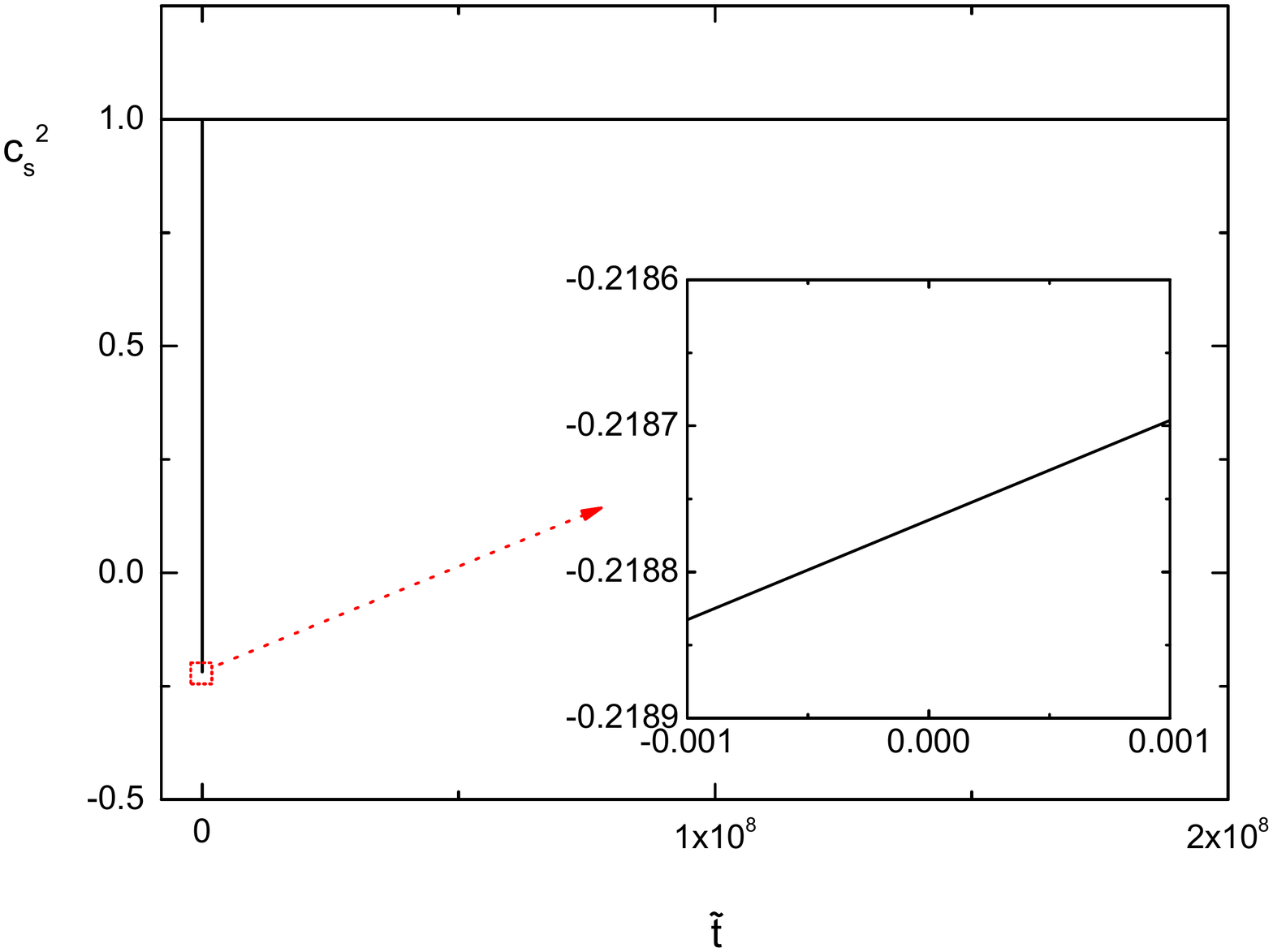}
% "\includegraphics" is very powerful; the graphicx package is already loaded
\caption{\label{fig:i} (Color plot.) The evolution of ${\cal G}$ and the sound speed square.
We have zoomed in the curves around the bounce point.}
\label{stbility}
\end{figure}

To be a successful model, there should not exist ghost instability or gradient instability
during the evolution. The ghost instability happens if the coefficient of time derivative ${\cal G}$
becomes negative, and the gradient instability happens if the sound speed square $c_s^2$ becomes negative.
Since both the formulae of ${\cal G}$ and $c_s^2$ are quite complicated to see their sign, we have done
numeric calculation and show the results in Figure \ref{stbility}. We find during the evolution,
${\cal G}$ is always positive, so our model is ghost-free. For $c_s^2$, it equals $+1$ in both
contracting phase and expanding phase, this can be read from Eq.(\ref{fff}) and Eq.(\ref{ggg}) when
both $\alpha$ and $\beta$ vanish, since we have used the delta function $c(h)$ to make them
nonzero only near the bounce point. However, when the universe bounce, $c_s^2$ becomes negative
(see the zoomed in plot at the down panel of Figure \ref{stbility}), so there is a brief gradient
instability near the bounce point. This kind of brief gradient instability also happen in similar
Galileon bounce models such as\cite{Koehn:2013upa,Qiu:2011cy,Qiu:2015nha}. There are lots of
discussions about it, in \cite{Koehn:2013upa} the authors point out this brief gradient instability
could be the hint to show the condition of $\delta h =0$ in the unitary gauge is violated near the
bounce zone.
%, in \cite{Qiu:2015nha} the authors introduce new terms to evade this instability.
Also, the model under consideration is not UV complete but phenomenologically constructed from the perspective of effective field viewpoint. This issue remains open in the study of nonsingular bounce cosmology by making use of field theory models.
%Here we won't go further, and regard it as a open question to be discussed carefully in future.

\section{Conclusions}
As the last found standard model particle and the only scalar particle, the Higgs boson
attracted more and more attention. Using the Higgs field, people are able to construct many
models to explain early universe physics, such as bounce inflation. The bounce inflation
scenario aims at avoiding the Standard Big-Bang Singularity by adding a nonsingular bounce
before the inflation period. Thus it requires NEC violation and Quintom-like matter. By
introducing a higher-derivative term into the Higgs action, it is possible to realize a
bounce. Moreover, an anisotropy-free bounce requires the EoS parameter to be larger than
unity in contracting phase, which is realized if the Higgs potential is negative in that
period. By taking into account loop-corrections introduced by the interactions of Higgs
with fermions such as the top quark, the Higgs effective potential can become negative.

In our model, this Higgs field starts from where the potential energy is negative, with a
leftward velocity in order to guarantee that the total energy density is positive. In the
contracting phase, the Higgs field will move leftwards and climb up along the potential.
Since the potential is negative, the EoS parameter in the contracting phase will be larger
than 1. Along with the evolution, the velocity gets larger and larger. At some point
when the G-term is triggered, it becomes possible to violate the NEC without ghost mode,
thus makes the universe bounce into an expanding phase. At the same time, the Higgs field
evolves to a position near the local maximum of the potential. After the bounce, the Higgs
field falls down along the potential to its false vacuum at the center, and the G-term
disappears due to the delta function like prefactor $c(h)$. In the expanding phase, the
evolution is quickly attracted to the solution with tiny velocity of the field, so the
field slowly rolls down the potential, driving inflation of nearly $60$ e-folds as in
usual Higgs inflation. After the field reaches the false vacuum at the center of the
potential, it becomes oscillating and reheats the universe.

Our numerical calculation has been divided into three phases, the contracting phase,
the bouncing phase and the expanding phase, they are connected by the matching condition
that $h$ and $h'$ are always continuous. A bounce model demands that at the bounce point
the total energy density equals zero and
the NEC is violated. Both of these conditions can be satisfied by the competition between
$\rho_k$ and a negative $\rho_e$ (also $p_k$ and $p_e$) in our model. Near the bounce point,
both $\rho_k$ and $\rho_e$ are much larger than $\rho_p$, that is to say the Higgs potential
is irrelevant to the existence of a cosmological bounce. We find that a $\lambda h^4$ potential
together with a Coleman-Weinberg one-loop effective potential as a inflationary model can not
generate the observed value for $n_s$ , so our model must be a oversimplified type of Higgs
Bounce Inflation model. In our future work, we must analyze the Higgs effective potential
thoroughly to use the Higgs Bounce Inflation model to explain the observational data.

We have not started cosmological perturbations in our model. However, as which obtained in
\cite{Piao:2003zm}, for the small scale perturbation modes, the scalar and tensor
power spectrum should be similar to the standard Higgs inflation models. Corrections will arise
on large scale modes, but as long as they don't exit the horizon in contracting phase, corrections
might only be some oscillating wiggles. For very large scale modes which exit the horizon before
the bounce, the spectrum will get blue-tilted, which might be responsible for the small-$l$
suppression observed in the CMB TT spectrum. So we can see both the fast-roll phase before
the slow-roll inflation, and the cosmological contraction phase, could shed light on this
small $l$ anomaly. We will investigate the detailed calculations of this model in our future
work.

\acknowledgments

We thank Prof. Mingzhe Li, Prof. Yunsong Piao, Prof. Junqing Xia, Dr. Siyu Li for useful
discussions. We also wish to thank Prof. Robert H. Brandenberger and Francis Duplessis
for useful comments on our manuscript. T.Q. thanks the hospitality of Institute of High
Energy Physic during his visit, which improves this work. Y.W, F.P.H. and X.Z. are
supported by the NSFC under grants Nos. 11121092, 11033005, 11375220 and also by the
CAS pilotB program. The work of T.Q. is supported by NSFC under Grant No: 11405069. YFC
is supported in part by the Chinese National Youth Thousand Talents Program and by the
USTC start-up funding(Grant No. KY2030000049). H.L. is supported in part by the NSFC
under Grant No. 11033005 and the youth innovation promotion association project and
the Outstanding young scientists project of the Chinese Academy of Sciences.



\begin{thebibliography}{99}


%\cite{Guth:1980zm}
\bibitem{Guth:1980zm}
  A.~H.~Guth,
  \emph{The Inflationary Universe: A Possible Solution to the Horizon and Flatness Problems},
  Phys.\ Rev.\ D {\bf 23}, 347 (1981).
  %%CITATION = PHRVA,D23,347;%%

%\cite{Linde:1981mu}
\bibitem{Linde:1981mu}
  A.~D.~Linde,
  \emph{A New Inflationary Universe Scenario: A Possible Solution of the Horizon, Flatness, Homogeneity, Isotropy and Primordial Monopole Problems},
  Phys.\ Lett.\ B {\bf 108}, 389 (1982).
  %%CITATION = PHLTA,B108,389;%%

%\cite{Albrecht:1982wi}
\bibitem{Albrecht:1982wi}
  A.~Albrecht and P.~J.~Steinhardt,
  \emph{Cosmology for Grand Unified Theories with Radiatively Induced Symmetry Breaking},
  Phys.\ Rev.\ Lett.\  {\bf 48}, 1220 (1982).
  %%CITATION = PRLTA,48,1220;%%

  %\cite{Starobinsky:1980te}
\bibitem{Starobinsky:1980te}
  A.~A.~Starobinsky,
  \emph{A New Type of Isotropic Cosmological Models Without Singularity},
  Phys.\ Lett.\ B {\bf 91}, 99 (1980).
  %%CITATION = PHLTA,B91,99;%%

%\cite{Fang:1980wi}
\bibitem{Fang:1980wi}
  L.~Z.~Fang,
  \emph{Entropy Generation In The Early Universe By Dissipative Processes Near The Higgs' Phase Transitions},
  Phys.\ Lett.\ B {\bf 95}, 154 (1980).
  %%CITATION = PHLTA,B95,154;%%

  %\cite{Sato:1980yn}
\bibitem{Sato:1980yn}
  K.~Sato,
  \emph{First Order Phase Transition of a Vacuum and Expansion of the Universe},
  Mon.\ Not.\ Roy.\ Astron.\ Soc.\  {\bf 195}, 467 (1981).
  %%CITATION = MNRAA,195,467;%%

\bibitem{Hawking1970}
  S.~W.~Hawking, and R.~Penrose,
  \emph{The Singularities of Gravitational Collapse and Cosmology},
  Pro.\ Roy.\ Soc.\ Lond.\ A
  {\bf 314}, 529 (1970).

\bibitem{Hawking1973}
  S.~W.~Hawking, and G.~F.~R.~Ellis,
  \emph{The Large Scale Structure of Space-Time},
  Cambridge, 1973.

\bibitem{Vilenkin1994a}
  Arvind Borde and Alexander Vilenkin,
  \emph{Eternal inflation and the initial singularity},
  Phys.\ Rev.\ Lett. {\bf 72}, 3305 (1994).


\bibitem{Vilenkin1994b}
  Arvind Borde, Alan H. Guth, and Alexander Vilenkin,
  \emph{Inflationary Spacetimes Are Incomplete in Past Directions},
   Phys.\ Rev.\ Lett. {\bf 90}, 151301 (1994).

 %\cite{Novello:2008ra}
\bibitem{Novello:2008ra}
  M.~Novello and S.~E.~P.~Bergliaffa,
  \emph{Bouncing Cosmologies},
  Phys.\ Rept.\  {\bf 463}, 127 (2008)
  [arXiv:0802.1634 [astro-ph]].
  %%CITATION = ARXIV:0802.1634;%%

\bibitem{Cai:2014bea}
  Y.~F.~Cai,
  \emph{Exploring Bouncing Cosmologies with Cosmological Surveys},
  Sci.\ China Phys.\ Mech.\ Astron.\  {\bf 57}, 1414 (2014)
  [arXiv:1405.1369 [hep-th]].
  %%CITATION = ARXIV:1405.1369;%%

\bibitem{Battefeld}
  D.~Battefeld and P.~Peter,
  \emph{A Critical Review of Classical Bouncing Cosmologies},
  Phys.\ Rept.\  {\bf 571}, 1 (2015)
  [arXiv:1406.2790 [astro-ph.CO]].

%\cite{Ayon-Beato:2015eca}
\bibitem{Ayon-Beato:2015eca}
  E.~Ayon-Beato, F.~Canfora and J.~Zanelli,
  \emph{Analytic self-gravitating Skyrmions, cosmological bounces and wormholes},
  arXiv:1509.02659 [gr-qc].
  %%CITATION = ARXIV:1509.02659;%%

\bibitem{Cai:2007qw}
  Y.~F.~Cai, T.~Qiu, Y.~S.~Piao, M.~Li and X.~Zhang,
  \emph{Bouncing universe with quintom matter},
  JHEP {\bf 0710}, 071 (2007)
  [arXiv:0704.1090 [gr-qc]].

%\cite{Feng:2004ad}
\bibitem{Feng:2004ad}
  B.~Feng, X.~L.~Wang and X.~M.~Zhang,
  \emph{Dark energy constraints from the cosmic age and supernova},
  Phys.\ Lett.\ B {\bf 607}, 35 (2005)
  [astro-ph/0404224].
  %%CITATION = ASTRO-PH/0404224;%%
  %824 citations counted in INSPIRE as of 31 Aug 2015

%\cite{Guo:2004fq}
\bibitem{Guo:2004fq}
  Z.~K.~Guo, Y.~S.~Piao, X.~M.~Zhang and Y.~Z.~Zhang,
  \emph{Cosmological evolution of a quintom model of dark energy},
  Phys.\ Lett.\ B {\bf 608}, 177 (2005)
  [astro-ph/0410654].
  %%CITATION = ASTRO-PH/0410654;%%
  %499 citations counted in INSPIRE as of 31 Aug 2015

%\cite{Zhang:2005eg}
\bibitem{Zhang:2005eg}
  X.~F.~Zhang, H.~Li, Y.~S.~Piao and X.~M.~Zhang,
  \emph{Two-field models of dark energy with equation of state across -1},
  Mod.\ Phys.\ Lett.\ A {\bf 21}, 231 (2006)
  [astro-ph/0501652].
  %%CITATION = ASTRO-PH/0501652;%%
  %194 citations counted in INSPIRE as of 31 Aug 2015

%\cite{Li:2005fm}
\bibitem{Li:2005fm}
  M.~Z.~Li, B.~Feng and X.~M.~Zhang,
  \emph{A Single scalar field model of dark energy with equation of state crossing -1},
  JCAP {\bf 0512}, 002 (2005)
  [hep-ph/0503268].
  %%CITATION = HEP-PH/0503268;%%
  %211 citations counted in INSPIRE as of 31 Aug 2015

\bibitem{Zhang:2006ck}
  X.~F.~Zhang and T.~Qiu,
  \emph{Avoiding the big-rip jeopardy in a quintom dark energy model with higher derivatives},
  Phys.\ Lett.\ B {\bf 642}, 187 (2006)
  [astro-ph/0603824].

\bibitem{Cai:2007gs}
  Y.~F.~Cai, M.~Z.~Li, J.~X.~Lu, Y.~S.~Piao, T.~T.~Qiu and X.~M.~Zhang,
  \emph{A String-Inspired Quintom Model Of Dark Energy},
  Phys.\ Lett.\ B {\bf 651}, 1 (2007)
  [hep-th/0701016].

\bibitem{Cai:2007zv}
  Y.~F.~Cai, T.~Qiu, R.~Brandenberger, Y.~S.~Piao and X.~Zhang,
  \emph{On Perturbations of Quintom Bounce},
  JCAP {\bf 0803}, 013 (2008)
  [arXiv:0711.2187 [hep-th]].
  %%CITATION = ARXIV:0711.2187;%%
  %102 citations counted in INSPIRE as of 31 Aug 2015

\bibitem{Cai:2008qb}
  Y.~F.~Cai, T.~T.~Qiu, J.~Q.~Xia and X.~Zhang,
  \emph{A Model Of Inflationary Cosmology Without Singularity},
  Phys.\ Rev.\ D {\bf 79}, 021303 (2009)
  [arXiv:0808.0819 [astro-ph]].

\bibitem{Cai:2008qw}
  Y.~F.~Cai, T.~T.~Qiu, R.~Brandenberger and X.~M.~Zhang,
  \emph{A Nonsingular Cosmology with a Scale-Invariant Spectrum of Cosmological Perturbations from Lee-Wick Theory},
  Phys.\ Rev.\ D {\bf 80}, 023511 (2009)
  [arXiv:0810.4677 [hep-th]].
  %%CITATION = ARXIV:0810.4677;%%
  %118 citations counted in INSPIRE as of 31 Aug 2015

\bibitem{Qiu:2010dk}
  T.~Qiu and K.~C.~Yang,
  \emph{Perturbations in Matter Bounce with Non-minimal Coupling},
  JCAP {\bf 1011}, 012 (2010)
  [arXiv:1007.2571 [astro-ph.CO]].

%\cite{Qiu:2011cy}
\bibitem{Qiu:2011cy}
  T.~Qiu, J.~Evslin, Y.~F.~Cai, M.~Li and X.~Zhang,
  \emph{Bouncing Galileon Cosmologies},
  JCAP {\bf 1110}, 036 (2011)
  [arXiv:1108.0593 [hep-th]].
  %%CITATION = ARXIV:1108.0593;%%
  %75 citations counted in INSPIRE as of 31 Aug 2015

\bibitem{Easson:2011zy}
  D.~A.~Easson, I.~Sawicki and A.~Vikman,
  \emph{G-Bounce},
  JCAP {\bf 1111}, 021 (2011)
  [arXiv:1109.1047 [hep-th]].

\bibitem{Cai:2012va}
  Y.~F.~Cai, D.~A.~Easson and R.~Brandenberger,
  \emph{Towards a Nonsingular Bouncing Cosmology},
  JCAP {\bf 1208}, 020 (2012)
  [arXiv:1206.2382 [hep-th]].

%\cite{Qiu:2013eoa}
\bibitem{Qiu:2013eoa}
  T.~Qiu, X.~Gao and E.~N.~Saridakis,
  \emph{Towards anisotropy-free and nonsingular bounce cosmology with scale-invariant perturbations},
  Phys.\ Rev.\ D {\bf 88}, No. 4, 043525 (2013)
  [arXiv:1303.2372 [astro-ph.CO]].
  %%CITATION = ARXIV:1303.2372;%%
  %43 citations counted in INSPIRE as of 31 Aug 2015

\bibitem{Koehn:2013upa}
  M.~Koehn, J.~L.~Lehners and B.~A.~Ovrut,
  \emph{Cosmological super-bounce},
  Phys.\ Rev.\ D {\bf 90}, No. 2, 025005 (2014)
  [arXiv:1310.7577 [hep-th]].

\bibitem{Battarra}
  L.~Battarra, M.~Koehn, J.~L.~Lehners and B.~A.~Ovrut,
  \emph{Cosmological Perturbations Through a Non-Singular Ghost-Condensate/Galileon Bounce},
  JCAP {\bf 1407}, 007 (2014)
  [arXiv:1404.5067 [hep-th]].

%\cite{Qiu:2015nha}
\bibitem{Qiu:2015nha}
  T.~Qiu and Y.~T.~Wang,
  \emph{G-Bounce Inflation: Towards Nonsingular Inflation Cosmology with Galileon Field},
  JHEP {\bf 1504}, 130 (2015)
  [arXiv:1501.03568 [astro-ph.CO]].
  %%CITATION = ARXIV:1501.03568;%%

\bibitem{Qiu:2010vk}
  T.~Qiu,
  \emph{Can the Big Bang Singularity be avoided by a single scalar field?},
  Class.\ Quant.\ Grav.\  {\bf 27}, 215013 (2010)
  [arXiv:1007.2929 [hep-ph]].

 %\cite{Cai:2009zp}
\bibitem{Cai:2009zp}
  Y.~F.~Cai, E.~N.~Saridakis, M.~R.~Setare and J.~Q.~Xia,
  \emph{Quintom Cosmology: Theoretical implications and observations},
  Phys.\ Rept.\  {\bf 493}, 1 (2010)
  [arXiv:0909.2776 [hep-th]],\\
  %%CITATION = ARXIV:0909.2776;%%
  %310 citations counted in INSPIRE as of 31 Aug 2015

\bibitem{Qiu:2010ux}
  T.~Qiu,
  \emph{Theoretical Aspects of Quintom Models},
  Mod.\ Phys.\ Lett.\ A {\bf 25}, 909 (2010)
  [arXiv:1002.3971 [hep-th]].

%\cite{Piao:2003zm}
\bibitem{Piao:2003zm}
  Y.~S.~Piao, B.~Feng and X.~M.~Zhang,
  \emph{Suppressing CMB quadrupole with a bounce from contracting phase to inflation},
  Phys.\ Rev.\ D {\bf 69}, 103520 (2004)
  [hep-th/0310206].
  %%CITATION = HEP-TH/0310206;%%
  %98 citations counted in INSPIRE as of 31 Aug 2015

%\cite{Piao:2005ag}
\bibitem{Piao:2005ag}
  Y.~S.~Piao,
  \emph{A Possible explanation to low CMB quadrupole},
  Phys.\ Rev.\ D {\bf 71}, 087301 (2005)
  [astro-ph/0502343].
  %%CITATION = ASTRO-PH/0502343;%%
  %38 citations counted in INSPIRE as of 31 Aug 2015

%\cite{Piao:2003hh}
\bibitem{Piao:2003hh}
  Y.~S.~Piao, S.~Tsujikawa and X.~M.~Zhang,
  \emph{Inflation in string inspired cosmology and suppression of CMB low multipoles},
  Class.\ Quant.\ Grav.\  {\bf 21}, 4455 (2004)
  [hep-th/0312139].
  %%CITATION = HEP-TH/0312139;%%
  %44 citations counted in INSPIRE as of 31 Aug 2015

%\cite{Kunze:1999xp}
\bibitem{Kunze:1999xp}
  K.~E.~Kunze and R.~Durrer,
  \emph{Anisotropic 'hairs' in string cosmology},
  Class.\ Quant.\ Grav.\  {\bf 17}, 2597 (2000)
  [gr-qc/9912081].
  %%CITATION = GR-QC/9912081;%%
  %16 citations counted in INSPIRE as of 31 Aug 2015
%\cite{Erickson:2003zm}

\bibitem{Erickson:2003zm}
  J.~K.~Erickson, D.~H.~Wesley, P.~J.~Steinhardt and N.~Turok,
  \emph{Kasner and mixmaster behavior in universes with equation of state w >= 1},
  Phys.\ Rev.\ D {\bf 69}, 063514 (2004)
  [hep-th/0312009].
  %%CITATION = HEP-TH/0312009;%%
  %143 citations counted in INSPIRE as of 31 Aug 2015
%\cite{Xue:2010ux}

\bibitem{Xue:2010ux}
  B.~Xue and P.~J.~Steinhardt,
  \emph{Unstable growth of curvature perturbation in non-singular bouncing cosmologies},
  Phys.\ Rev.\ Lett.\  {\bf 105}, 261301 (2010)
  [arXiv:1007.2875 [hep-th]].
  %%CITATION = ARXIV:1007.2875;%%
  %30 citations counted in INSPIRE as of 31 Aug 2015
%\cite{Xue:2011nw}

\bibitem{Xue:2011nw}
  B.~Xue and P.~J.~Steinhardt,
  \emph{Evolution of curvature and anisotropy near a nonsingular bounce},
  Phys.\ Rev.\ D {\bf 84}, 083520 (2011)
  [arXiv:1106.1416 [hep-th]].
  %%CITATION = ARXIV:1106.1416;%%
  %38 citations counted in INSPIRE as of 31 Aug 2015


%\cite{Khoury:2001wf}
\bibitem{Khoury:2001wf}
  J.~Khoury, B.~A.~Ovrut, P.~J.~Steinhardt and N.~Turok,
  \emph{The Ekpyrotic universe: Colliding branes and the origin of the hot big bang},
  Phys.\ Rev.\ D {\bf 64}, 123522 (2001)
  [hep-th/0103239].
  %%CITATION = HEP-TH/0103239;%%

%\cite{Aad:2012tfa}
\bibitem{Aad:2012tfa}
  G.~Aad {\it et al.} [ATLAS Collaboration],
  \emph{Observation of a new particle in the search for the Standard Model Higgs boson with the ATLAS detector at the LHC},
  Phys.\ Lett.\ B {\bf 716}, 1 (2012)
  [arXiv:1207.7214 [hep-ex]].
  %%CITATION = ARXIV:1207.7214;%%
  %4879 citations counted in INSPIRE as of 31 Aug 2015

%\cite{Chatrchyan:2012xdj}
\bibitem{Chatrchyan:2012xdj}
  S.~Chatrchyan {\it et al.} [CMS Collaboration],
  \emph{Observation of a new boson at a mass of 125 GeV with the CMS experiment at the LHC},
  Phys.\ Lett.\ B {\bf 716}, 30 (2012)
  [arXiv:1207.7235 [hep-ex]].
  %%CITATION = ARXIV:1207.7235;%%
  %4782 citations counted in INSPIRE as of 31 Aug 2015


%\cite{Espinosa:2007qp}
\bibitem{Espinosa:2007qp}
  J.~R.~Espinosa, G.~F.~Giudice and A.~Riotto,
  \emph{Cosmological implications of the Higgs mass measurement},
  JCAP {\bf 0805}, 002 (2008)
  [arXiv:0710.2484 [hep-ph]].
  %%CITATION = ARXIV:0710.2484;%%
  %138 citations counted in INSPIRE as of 31 Aug 2015
%\cite{Bars:2013vba}

\bibitem{Bars:2013vba}
  I.~Bars, P.~J.~Steinhardt and N.~Turok,
  \emph{Cyclic Cosmology, Conformal Symmetry and the Metastability of the Higgs},
  Phys.\ Lett.\ B {\bf 726}, 50 (2013)
  [arXiv:1307.8106 [gr-qc]].
  %%CITATION = ARXIV:1307.8106;%%
  %29 citations counted in INSPIRE as of 31 Aug 2015

%\cite{CervantesCota:1995tz}
\bibitem{CervantesCota:1995tz}
  J.~L.~Cervantes-Cota and H.~Dehnen,
  \emph{Induced gravity inflation in the standard model of particle physics},
  Nucl.\ Phys.\ B {\bf 442}, 391 (1995)
  [astro-ph/9505069].
  %%CITATION = ASTRO-PH/9505069;%%

\bibitem{Bezrukov:2007ep}
  F.~L.~Bezrukov and M.~Shaposhnikov,
  \emph{The Standard Model Higgs boson as the inflaton},
  Phys.\ Lett.\ B {\bf 659}, 703 (2008)
  [arXiv:0710.3755 [hep-th]].
  %%CITATION = ARXIV:0710.3755;%%
  %630 citations counted in INSPIRE as of 31 Aug 2015
%\cite{Bezrukov:2013fka}

\bibitem{DeSimone:2008ei}
  A.~De Simone, M.~P.~Hertzberg and F.~Wilczek,
  \emph{Running Inflation in the Standard Model},
  Phys.\ Lett.\ B {\bf 678}, 1 (2009)
  [arXiv:0812.4946 [hep-ph]].

%\cite{Burgess:2009ea}
\bibitem{Burgess:2009ea}
  C.~P.~Burgess, H.~M.~Lee and M.~Trott,
  \emph{Power-counting and the Validity of the Classical Approximation During Inflation},
  JHEP {\bf 0909}, 103 (2009)
  [arXiv:0902.4465 [hep-ph]].
  %%CITATION = ARXIV:0902.4465;%%
%
%\cite{Barbon:2009ya}
\bibitem{Barbon:2009ya}
  J.~L.~F.~Barbon and J.~R.~Espinosa,
  \emph{On the Naturalness of Higgs Inflation},
  Phys.\ Rev.\ D {\bf 79}, 081302 (2009)
  [arXiv:0903.0355 [hep-ph]].
  %%CITATION = ARXIV:0903.0355;%%

%\cite{Barvinsky:2009fy}
\bibitem{Barvinsky:2009fy}
  A.~O.~Barvinsky, A.~Y.~.Kamenshchik, C.~Kiefer, A.~A.~Starobinsky and C.~Steinwachs,
  \emph{Asymptotic freedom in inflationary cosmology with a non-minimally coupled Higgs field},
  JCAP {\bf 0912}, 003 (2009)
  [arXiv:0904.1698 [hep-ph]].
  %%CITATION = ARXIV:0904.1698;%%
%
%\cite{Lerner:2009na}
\bibitem{Lerner:2009na}
  R.~N.~Lerner and J.~McDonald,
  \emph{Higgs Inflation and Naturalness},
  JCAP {\bf 1004}, 015 (2010)
  [arXiv:0912.5463 [hep-ph]].
  %%CITATION = ARXIV:0912.5463;%%
%
%\cite{Burgess:2010zq}
\bibitem{Burgess:2010zq}
  C.~P.~Burgess, H.~M.~Lee and M.~Trott,
  \emph{Comment on Higgs Inflation and Naturalness},
  JHEP {\bf 1007}, 007 (2010)
  [arXiv:1002.2730 [hep-ph]].
  %%CITATION = ARXIV:1002.2730;%%

%%\cite{Hertzberg:2010dc}
\bibitem{Hertzberg:2010dc}
  M.~P.~Hertzberg,
  \emph{On Inflation with Non-minimal Coupling},
  JHEP {\bf 1011}, 023 (2010)
  [arXiv:1002.2995 [hep-ph]].
  %%CITATION = ARXIV:1002.2995;%%
  %120 citations counted in INSPIRE as of 31 Aug 2015

%\cite{Germani:2010gm}
\bibitem{Germani:2010gm}
  C.~Germani and A.~Kehagias,
  \emph{New Model of Inflation with Non-minimal Derivative Coupling of Standard Model Higgs Boson to Gravity},
  Phys.\ Rev.\ Lett.\  {\bf 105}, 011302 (2010)
  [arXiv:1003.2635 [hep-ph]].
  %%CITATION = ARXIV:1003.2635;%%
  %166 citations counted in INSPIRE as of 31 Aug 2015

%\cite{Bezrukov:2010jz}
\bibitem{Bezrukov:2010jz}
  F.~Bezrukov, A.~Magnin, M.~Shaposhnikov and S.~Sibiryakov,
  \emph{Higgs inflation: consistency and generalisations},
  JHEP {\bf 1101}, 016 (2011)
  [arXiv:1008.5157 [hep-ph]].
  %%CITATION = ARXIV:1008.5157;%%
%
%\cite{Atkins:2010yg}
\bibitem{Atkins:2010yg}
  M.~Atkins and X.~Calmet,
  \emph{Remarks on Higgs Inflation},
  Phys.\ Lett.\ B {\bf 697}, 37 (2011)
  [arXiv:1011.4179 [hep-ph]].
  %%CITATION = ARXIV:1011.4179;%%
%

%\cite{Kamada:2010qe}
\bibitem{Kamada:2010qe}
  K.~Kamada, T.~Kobayashi, M.~Yamaguchi and J.~Yokoyama,
  \emph{Higgs G-inflation},
  Phys.\ Rev.\ D {\bf 83}, 083515 (2011)
  [arXiv:1012.4238 [astro-ph.CO]].
  %%CITATION = ARXIV:1012.4238;%%
  %86 citations counted in INSPIRE as of 31 Aug 2015

%\cite{Horvat:2011wr}
\bibitem{Horvat:2011wr}
  R.~Horvat,
  \emph{Holographic bounds and Higgs inflation},
  Phys.\ Lett.\ B {\bf 699}, 174 (2011)
  [arXiv:1101.0721 [hep-ph]].
  %%CITATION = ARXIV:1101.0721;%%

\bibitem{Qiu:2011tk}
  T.~Qiu and D.~Maity,
  \emph{Higgs Inflation in Horava-Lifshitz Gravity},
  arXiv:1104.4386 [hep-th].

\bibitem{Bezrukov:2013fka}
  F.~Bezrukov,
  \emph{The Higgs field as an inflaton},
  Class.\ Quant.\ Grav.\  {\bf 30}, 214001 (2013)
  [arXiv:1307.0708 [hep-ph]].
  %%CITATION = ARXIV:1307.0708;%%
  %47 citations counted in INSPIRE as of 31 Aug 2015

%\cite{Kearney:2015vba}
\bibitem{Kearney:2015vba}
  J.~Kearney, H.~Yoo and K.~M.~Zurek,
  \emph{Is a Higgs Vacuum Instability Fatal for High-Scale Inflation?},
  Phys.\ Rev.\ D {\bf 91}, No. 12, 123537 (2015)
  [arXiv:1503.05193 [hep-th]].
  %%CITATION = ARXIV:1503.05193;%%
  %7 citations counted in INSPIRE as of 31 Aug 2015

  %\cite{Moss:2015fma}
\bibitem{Moss:2015fma}
  I.~G.~Moss,
  \emph{Higgs boson cosmology},
  arXiv:1507.05760 [hep-ph].
  %%CITATION = ARXIV:1507.05760;%%


  %\cite{Huang:2013oua}
\bibitem{Huang:2013oua}
  F.~P.~Huang, C.~S.~Li, D.~Y.~Shao and J.~Wang,
  \emph{Phenomenology of an Extended Higgs Portal Inflation Model after Planck 2013},
  Eur.\ Phys.\ J.\ C {\bf 74}, No. 8, 2990 (2014)
  [arXiv:1307.7458 [hep-ph]].
  %%CITATION = ARXIV:1307.7458;%%
  %4 citations counted in INSPIRE as of 05 Oct 2015



  %\cite{Hamada:2014iga}
\bibitem{Hamada:2014iga}
  Y.~Hamada, H.~Kawai, K.~Y.~Oda and S.~C.~Park,
  \emph{Higgs Inflation is Still Alive after the Results from BICEP2},
  Phys.\ Rev.\ Lett.\  {\bf 112}, No. 24, 241301 (2014)
  [arXiv:1403.5043 [hep-ph]].
  %%CITATION = ARXIV:1403.5043;%%
  %79 citations counted in INSPIRE as of 05 Oct 2015

  %\cite{Hamada:2014wna}
\bibitem{Hamada:2014wna}
  Y.~Hamada, H.~Kawai, K.~Y.~Oda and S.~C.~Park,
  \emph{Higgs inflation from Standard Model criticality},
  Phys.\ Rev.\ D {\bf 91}, 053008 (2015)
  [arXiv:1408.4864 [hep-ph]].
  %%CITATION = ARXIV:1408.4864;%%
  %26 citations counted in INSPIRE as of 05 Oct 2015


%\cite{Salvio:2013rja}
\bibitem{Salvio:2013rja}
  A.~Salvio,
  %``Higgs Inflation at NNLO after the Boson Discovery,''
  Phys.\ Lett.\ B {\bf 727}, 234 (2013)
  [arXiv:1308.2244 [hep-ph]].
  %%CITATION = ARXIV:1308.2244;%%
  %32 citations counted in INSPIRE as of 15 Nov 2015

  %\cite{Salvio:2015kka}
\bibitem{Salvio:2015kka}
  A.~Salvio and A.~Mazumdar,
  %``Classical and Quantum Initial Conditions for Higgs Inflation,''
  Phys.\ Lett.\ B {\bf 750}, 194 (2015)
  [arXiv:1506.07520 [hep-ph]].
  %%CITATION = ARXIV:1506.07520;%%
  %3 citations counted in INSPIRE as of 15 Nov 2015


%\cite{Cai:2012qi}
\bibitem{Cai:2012qi}
  Y.~-F.~Cai and D.~A.~Easson,
  \emph{Higgs Boson in RG running Inflationary Cosmology},
  arXiv:1202.1285 [hep-th].
  %%CITATION = ARXIV:1202.1285;%%

%\cite{Kunimitsu:2012xx}
\bibitem{Kunimitsu:2012xx}
  T.~Kunimitsu and J.~Yokoyama,
  \emph{Higgs condensation as an unwanted curvaton},
  Phys.\ Rev.\ D {\bf 86}, 083541 (2012)
  [arXiv:1208.2316 [hep-ph]].
  %%CITATION = ARXIV:1208.2316;%%

%\cite{Choi:2012cp}
\bibitem{Choi:2012cp}
  K.~-Y.~Choi and Q.~-G.~Huang,
  \emph{Can Standard Model Higgs Seed the Formation of Structures in Our Universe?},
  arXiv:1209.2277 [hep-ph].
  %%CITATION = ARXIV:1209.2277;%%

%\cite{DeSimone:2012qr}
\bibitem{DeSimone:2012qr}
  A.~De Simone and A.~Riotto,
  \emph{Cosmological Perturbations from the Standard Model Higgs},
  arXiv:1208.1344 [hep-ph].
  %%CITATION = ARXIV:1208.1344;%%

%\cite{DeSimone:2012gq}
\bibitem{DeSimone:2012gq}
  A.~De Simone, H.~Perrier and A.~Riotto,
  \emph{Non-Gaussianities from the Standard Model Higgs},
  arXiv:1210.6618 [hep-ph].
  %%CITATION = ARXIV:1210.6618;%%

\bibitem{Cai:2013caa}
  Y.~F.~Cai, Y.~C.~Chang, P.~Chen, D.~A.~Easson and T.~Qiu,
  \emph{Planck constraints on Higgs modulated reheating of renormalization group improved inflation},
  Phys.\ Rev.\ D {\bf 88}, 083508 (2013)
  [arXiv:1304.6938 [hep-th]].





%\cite{Buttazzo:2013uya}
\bibitem{Buttazzo:2013uya}
  D.~Buttazzo, G.~Degrassi, P.~P.~Giardino, G.~F.~Giudice, F.~Sala, A.~Salvio and A.~Strumia,
  \emph{Investigating the near-criticality of the Higgs boson},
  JHEP {\bf 1312}, 089 (2013)
  [arXiv:1307.3536 [hep-ph]].
  %%CITATION = ARXIV:1307.3536;%%
  %339 citations counted in INSPIRE as of 31 Aug 2015


%\cite{Degrassi:2012ry}
\bibitem{Degrassi:2012ry}
  G.~Degrassi, S.~Di Vita, J.~Elias-Miro, J.~R.~Espinosa, G.~F.~Giudice, G.~Isidori and A.~Strumia,
  \emph{Higgs mass and vacuum stability in the Standard Model at NNLO},
  JHEP {\bf 1208}, 098 (2012)
  [arXiv:1205.6497 [hep-ph]].
  %%CITATION = ARXIV:1205.6497;%%
  %514 citations counted in INSPIRE as of 31 Aug 2015







%\cite{Nicolis:2008in}
\bibitem{Nicolis:2008in}
  A.~Nicolis, R.~Rattazzi and E.~Trincherini,
  \emph{The Galileon as a local modification of gravity},
  Phys.\ Rev.\ D {\bf 79}, 064036 (2009)
  [arXiv:0811.2197 [hep-th]].
  %%CITATION = ARXIV:0811.2197;%%
  %712 citations counted in INSPIRE as of 31 Aug 2015

%\cite{Deffayet:2009wt}
\bibitem{Deffayet:2009wt}
  C.~Deffayet, G.~Esposito-Farese and A.~Vikman,
  \emph{Covariant Galileon},
  Phys.\ Rev.\ D {\bf 79}, 084003 (2009)
  [arXiv:0901.1314 [hep-th]].
  %%CITATION = ARXIV:0901.1314;%%
  %384 citations counted in INSPIRE as of 31 Aug 2015

%\cite{Nicolis:2009qm}
\bibitem{Nicolis:2009qm}
  A.~Nicolis, R.~Rattazzi and E.~Trincherini,
  \emph{Energy's and amplitudes' positivity},
  JHEP {\bf 1005}, 095 (2010)
  [JHEP {\bf 1111}, 128 (2011)]
  [arXiv:0912.4258 [hep-th]].
  %%CITATION = ARXIV:0912.4258;%%
  %74 citations counted in INSPIRE as of 31 Aug 2015

%\cite{Deffayet:2009mn}
\bibitem{Deffayet:2009mn}
  C.~Deffayet, S.~Deser and G.~Esposito-Farese,
  \emph{Generalized Galileons: All scalar models whose curved background extensions maintain second-order field equations and stress-tensors},
  Phys.\ Rev.\ D {\bf 80}, 064015 (2009)
  [arXiv:0906.1967 [gr-qc]].
  %%CITATION = ARXIV:0906.1967;%%
  %292 citations counted in INSPIRE as of 31 Aug 2015

%\cite{Deffayet:2011gz}
\bibitem{Deffayet:2011gz}
  C.~Deffayet, X.~Gao, D.~A.~Steer and G.~Zahariade,
  \emph{From k-essence to generalised Galileons},
  Phys.\ Rev.\ D {\bf 84}, 064039 (2011)
  [arXiv:1103.3260 [hep-th]].
  %%CITATION = ARXIV:1103.3260;%%
  %264 citations counted in INSPIRE as of 31 Aug 2015


%\cite{Brandenberger:2015nua}
\bibitem{Brandenberger:2015nua}
  R.~H.~Brandenberger, Y.~F.~Cai, Y.~Wan and X.~Zhang,
  \emph{Nonsingular Cosmology from an Unstable Higgs Field},
  arXiv:1506.06770 [hep-th].
  %%CITATION = ARXIV:1506.06770;%%
  %1 citations counted in INSPIRE as of 31 Aug 2015



  %\cite{Cai:2013kja}
\bibitem{Cai:2013kja}
  Y.~F.~Cai, E.~McDonough, F.~Duplessis and R.~H.~Brandenberger,
  \emph{Two Field Matter Bounce Cosmology},
  JCAP {\bf 1310}, 024 (2013)
  [arXiv:1305.5259 [hep-th]].
  %%CITATION = ARXIV:1305.5259;%%
  %40 citations counted in INSPIRE as of 05 Oct 2015



%\cite{Finelli:2001sr}
\bibitem{Finelli:2001sr}
  F.~Finelli and R.~Brandenberger,
  \emph{On the generation of a scale invariant spectrum of adiabatic fluctuations in cosmological models with a contracting phase},
  Phys.\ Rev.\ D {\bf 65}, 103522 (2002)
  [hep-th/0112249].
  %%CITATION = HEP-TH/0112249;%%
  %203 citations counted in INSPIRE as of 31 Aug 2015

%\cite{Wands:1998yp}
\bibitem{Wands:1998yp}
  D.~Wands,
  \emph{Duality invariance of cosmological perturbation spectra},
  Phys.\ Rev.\ D {\bf 60}, 023507 (1999)
  [gr-qc/9809062].
  %%CITATION = GR-QC/9809062;%%
  %162 citations counted in INSPIRE as of 31 Aug 2015




\bibitem{ColemanWeinberg}
  Sidney Coleman and Erick Weinberg,
  \emph{Radiative Corrections as the Origin of Spontaneous Symmetry Breaking},
  Phys.\ Rev .\ D {\bf 7}, 1888 (1973)



\bibitem{Sher89}
  M.~Sher,
  \emph{Electroweak higgs potentials and vacuum stability},
  Phys.\ Rept.\  {\bf 179}, 273 (1989).

\bibitem{RHBRMP}

  R.~H.~Brandenberger,
  \emph{Quantum Field Theory Methods and Inflationary Universe Models},
  Rev.\ Mod.\ Phys.\  {\bf 57}, 1 (1985).






%\cite{ArkaniHamed:2008ym}
\bibitem{ArkaniHamed:2008ym}
  N.~Arkani-Hamed, S.~Dubovsky, L.~Senatore and G.~Villadoro,
  \emph{(No) Eternal Inflation and Precision Higgs Physics},
  JHEP {\bf 0803}, 075 (2008)
  [arXiv:0801.2399 [hep-ph]].
  %%CITATION = ARXIV:0801.2399;%%
  %43 citations counted in INSPIRE as of 31 Aug 2015



\bibitem{Datta:1999dw}
  A.~Datta and X.~Zhang,
  \emph{Vacuum stability Higgs mass bound revisited with implications for extra dimension theories},
  Phys.\ Rev.\ D {\bf 61}, 074033 (2000)
  [hep-ph/9912450].


\bibitem{Mukhanov:1990me}
  V.~F.~Mukhanov, H.~A.~Feldman and R.~H.~Brandenberger,
  \emph{Theory of cosmological perturbations.},
  Phys.\ Rept.\  {\bf 215}, 203 (1992).

%\cite{Ade:2015lrj}
\bibitem{Ade:2015lrj}
  P.~A.~R.~Ade {\it et al.} [Planck Collaboration],
  \emph{Planck 2015 results. XX. Constraints on inflation},
  arXiv:1502.02114 [astro-ph.CO].
  %%CITATION = ARXIV:1502.02114;%%
  %257 citations counted in INSPIRE as of 31 Aug 2015

%\cite{Hinshaw:2012aka}
\bibitem{Hinshaw:2012aka}
  G.~Hinshaw {\it et al.} [WMAP Collaboration],
  \emph{Nine-Year Wilkinson Microwave Anisotropy Probe (WMAP) Observations: Cosmological Parameter Results},
  Astrophys.\ J.\ Suppl.\  {\bf 208}, 19 (2013)
  [arXiv:1212.5226 [astro-ph.CO]].
  %%CITATION = ARXIV:1212.5226;%%
  %1657 citations counted in INSPIRE as of 31 Aug 2015



%\cite{Contaldi:2003zv}
\bibitem{Contaldi:2003zv}
  C.~R.~Contaldi, M.~Peloso, L.~Kofman and A.~D.~Linde,
  \emph{Suppressing the lower multipoles in the CMB anisotropies},
  JCAP {\bf 0307}, 002 (2003)
  [astro-ph/0303636].
  %%CITATION = ASTRO-PH/0303636;%%
  %238 citations counted in INSPIRE as of 19 sept. 2015


%\cite{Feng:2003zua}
\bibitem{Feng:2003zua}
  B.~Feng and X.~Zhang,
  \emph{Double inflation and the low cmb quadrupole},
  Phys.\ Lett.\ B {\bf 570}, 145 (2003)
  [astro-ph/0305020].
  %%CITATION = ASTRO-PH/0305020;%%
  %105 citations counted in INSPIRE as of 31 Aug 2015

%\cite{Liu:2010fm}
\bibitem{Liu:2010fm}
  J.~Liu, Y.~F.~Cai and H.~Li,
  %``Evidences for bouncing evolution before inflation in cosmological surveys,''
  J.\ Theor.\ Phys.\  {\bf 1}, 1 (2012)
  [arXiv:1009.3372 [astro-ph.CO]].
  %%CITATION = ARXIV:1009.3372;%%

%\cite{Liu:2013kea}
\bibitem{Liu:2013kea}
  Z.~G.~Liu, Z.~K.~Guo and Y.~S.~Piao,
  \emph{Obtaining the CMB anomalies with a bounce from the contracting phase to inflation},
  Phys.\ Rev.\ D {\bf 88}, 063539 (2013)
  [arXiv:1304.6527 [astro-ph.CO]].
  %%CITATION = ARXIV:1304.6527;%%
  %64 citations counted in INSPIRE as of 31 Aug 2015

%\cite{Xia:2014tda}
\bibitem{Xia:2014tda}
  J.~Q.~Xia, Y.~F.~Cai, H.~Li and X.~Zhang,
  \emph{Evidence for bouncing evolution before inflation after BICEP2},
  Phys.\ Rev.\ Lett.\  {\bf 112}, 251301 (2014)
  [arXiv:1403.7623 [astro-ph.CO]].
  %%CITATION = ARXIV:1403.7623;%%
  %32 citations counted in INSPIRE as of 31 Aug 2015

%\cite{Wan:2014fra}
\bibitem{Wan:2014fra}
  Y.~Wan, S.~Li, M.~Li, T.~Qiu, Y.~Cai and X.~Zhang,
  \emph{Single field inflation with modulated potential in light of the Planck and BICEP2},
  Phys.\ Rev.\ D {\bf 90}, No. 2, 023537 (2014)
  [arXiv:1405.2784 [astro-ph.CO]].
  %%CITATION = ARXIV:1405.2784;%%
  %18 citations counted in INSPIRE as of 31 Aug 2015

  %\cite{Kobayashi:2010cm}
\bibitem{Kobayashi:2010cm}
  T.~Kobayashi, M.~Yamaguchi and J.~Yokoyama,
  %``G-inflation: Inflation driven by the Galileon field,''
  Phys.\ Rev.\ Lett.\  {\bf 105}, 231302 (2010)
  [arXiv:1008.0603 [hep-th]].
  %%CITATION = ARXIV:1008.0603;%%
  %174 citations counted in INSPIRE as of 08 Nov 2015






% Please avoid comments such as "For a review'', "For some examples",
% "and references therein" or move them in the text. In general,
% please leave only references in the bibliography and move all
% accessory text in footnotes.

% Also, please have only one work for each \bibitem.


\end{thebibliography}
\end{document}